\documentclass[aps,twocolumn,amsmath,amssymb,preprintnumbers]{revtex4}
\usepackage{amsmath} \usepackage{amsfonts} \usepackage{amssymb}
\usepackage{bbm}
\usepackage{epsfig}
\usepackage{graphics}
\usepackage{graphicx}
\newcommand{\be}{\begin{equation}}
\newcommand{\ee}{\end{equation}}
\newcommand{\ba}{\begin{eqnarray}}
\newcommand{\ea}{\end{eqnarray}}
\newcommand{\nn}{\nonumber}
\newcommand{\kr}{\rangle}
\newcommand{\kl}{\langle}

\newcommand{\cD}{{\cal D}}
\newcommand{\cF}{{\cal F}}
\newcommand{\tcu}{\tilde{\cal U}}

\newcommand{\tn}{(t_{n+1},t_n)}
\newcommand{\n}{_{n_+,n_-}}

\newcommand{\bv}{\hat\varphi}
\newcommand{\tnn}{(t_{n+1})}
\newcommand{\cC}{{\cal C}}

\newcommand{\cK}{{\cal K}}
\newcommand{\tep}{t-\epsilon}
\newcommand{\tepi}{t_{in}-\epsilon}
\newcommand{\ti}{t_{in}}

\begin{document}

\title[ ]{Fermions from classical statistics}

\author{C. Wetterich}
\affiliation{Institut  f\"ur Theoretische Physik\\
Universit\"at Heidelberg\\
Philosophenweg 16, D-69120 Heidelberg}

\begin{abstract}
We describe fermions in terms of a classical statistical ensemble. The states $\tau$ of this ensemble are characterized by a sequence of values one or zero or a corresponding set of two-level observables. Every classical probability distribution can be associated to a quantum state for fermions. If the time evolution of the classical probabilities $p_\tau$ amounts to a rotation of the wave function $q_\tau(t)=\pm \sqrt{p_\tau(t)}$, we infer the unitary time evolution of a quantum system of fermions according to a Schr\"odinger equation. We establish how such classical statistical ensembles can be mapped to Grassmann functional integrals. Quantum field theories for fermions arise for a suitable time evolution of classical probabilities for generalized Ising models.
\end{abstract}

\maketitle

\section{Introduction}
\label{Introduction}
It is widely believed that fermions are a genuine quantum concept that is not present in classical statistics. In contrast, we establish here that fermions arise rather naturally from a classical statistical ensemble. This is not in contradiction with  the quantum features of fermions. Actually, all the usual quantum features arise together with the fermions from suitable classical statistical ensembles. We encounter here an example for the emergence of quantum physics from classical statistics \cite{GR}, \cite{3A}.

A classical statistical ensemble is characterized by the states $\tau$ and the normalized positive probabilities $p_\tau$ for every state, $p_\tau\ge0$, $\sum_{\tau}p_\tau=1$. We consider here a set of two-level observables $N_\alpha$ which take in every state $\tau$ the values zero or one. We further assume that the states can be completely labeled by these values $n_\alpha=0,1$. This means that $\tau=[n_\beta]$ can be specified by an ordered set of numbers $0$ or $1$ for the $B$ two-level observables, $\alpha, \beta=1\dots B$. For finite $B$ there are $2^B$ states, and we also consider the limit $B\to\infty$. This setting is a very general one, since arbitrary observables can be composed from two-level observables. For example, the position observable of a particle can be defined by dividing space into cells by yes/no questions as upper half or lower half, left or right etc., and making such a grid as fine as required by a given experiment. Each yes/no question can be associated with a two-level observable. Already at this early stage the possible association between the set of numbers $[n_\beta]$ and the occupation numbers for a quantum system of fermions is suggestive.

If $\alpha$ consists of a position label $x$, for example denoting the points of a $D$-dimensional hypercubic lattice, plus an ``internal index'' $\gamma=1\dots f,~\alpha=(x,\gamma)$, the states $\tau$ are the possible states of a classical Ising model with $f$ different species. Instead of the occupation numbers $n_\alpha$ we can also use unit spins $S_\alpha=2n_\alpha-1=\pm 1$. In this case every probability distribution $\{p_\tau\}$ denotes a generalized Ising model. (A particular family of $\{p_\tau\}$ corresponds to the standard Ising model with next neighbor interactions.) We will associate to each $\{p_\tau\}$ a quantum state for a multi-fermion system. In this sense we map classical statistical ensembles of a generalized Ising type to quantum fermions. 

A crucial concept for the description of fermions is the ``classical wave function'' $\{q_\tau \}$, which consists of the positive or negative roots of the probabilities \cite{CWP}
\be \label{AA1}
p_\tau=q_\tau^2.
\ee
As we have discussed earlier \cite{CWP}, \cite{3A}, the relative signs of the real numbers $q_\tau$ are largely fixed by continuity properties, such that only few relative signs without physical relevance remain free. If all physical quantities are ultimately computable in terms of the probability distribution $\{p_\tau\}$, the signs in the classical wave function $\{q_\tau\}$ carry no independent relevant information. The choice of these signs can be associated with a choice of gauge. We can then use $\{q_\tau\}$ or $\{p_\tau\}$ as equivalent descriptions of the classical statistical ensemble. For all observables and correlations built from $N_\alpha$ the sign $s_\tau=\pm 1$ of $q_\tau$,
\be \label{AA2}
q_\tau=s_\tau\sqrt{p_\tau},
\ee
obviously plays no role, since expectation values can be computed from $\{p_\tau\}$.

One may further introduce a large class of `` statistical observables'' \cite{CWP}. They can be represented as off-diagonal operators acting on $q_\tau$. For those the signs $s_\tau$ matter, but they are fixed in a way such that the statistical observables remain computable in terms of $\{p_\tau\}$. Again, for such observables all necessary information is contained in the probability distribution $\{p_\tau\}$. In the present paper the statistical observables will not play a central role and we rather concentrate on the ``diagonal'' observables built from $N_\alpha$ which take a fixed value $A_\tau$ for every classical state $\tau$. Nevertheless, the use of the classical wave function $\{q_\tau\}$ is very convenient for the description of the classical statistical ensemble and constitutes the key for its interpretation in terms of fermions.  (There are some common features between the classical wave function and the Hilbert space formulation of classical mechanics by Koopman and von Neumann \cite{Kop}. In contrast to this work the classical wave function is real, however, and this will be a crucial ingredient for our approach.)

In this paper we concentrate on the description of the ensemble in terms of the classical wave function. We will identify $\{q_\tau\}$ with a quantum wave function. In the case of Ising like systems with $\alpha=(x,\gamma)$ this wave function describes a quantum theory for an arbitrary number of fermions. We further map each wave function $\{q_\tau\}$ to an element of a Grassmann algebra. This allows the implementation of the fermionic creation and annihilation operators and the discussion of quantum states with a fixed number of fermions. We discuss symmetries that can be realized in a simple way on the level of fermions, while on the level of the classical probability distribution $\{p_\tau\}$ they are rather hidden in the form of non-linear transformations. All this discussion is independent of the particular dynamics of the system. The first part of our paper therefore constitutes a very general map from a discrete classical statistical ensemble to a quantum theory of fermions. In this respect it is in line with earlier implementations of fermions within bosonic systems \cite{FB}.

The second part of this paper concerns the time evolution of the probability distribution of the classical statistical ensemble. For this purpose we need an evolution law which describes how $\{p_\tau\}$ depends on time $t$. This evolution law has to be specified as the basic dynamical law - it is not known a priori. There is substantial freedom in the choice of such a law - one only has to guarantee that all probabilities stay positive at all time, $p_\tau(t)\geq 0$, and that the normalization is preserved, $\sum_\tau p_\tau(t)=0$. 

We will discuss a specific form of the time evolution of the probability distribution as given by the ``unitary evolution law''
\be\label{A}
q_\tau (t)=\sum_\rho R_{\tau\rho}q_\rho(t-\epsilon),
\ee
with rotation matrix $R$ independent of $\{q_\tau\}$. (The most general evolution of $\{p_\tau\}$ can be described with $R$ depending on $\{q_\tau\}$.) The evolution law allows the computation of the classical wave function $q_\tau(t)$ at a time $t$ in terms of the wave function $q_\tau(t-\epsilon)$ at a previous time $t-\epsilon$ which may be infinitesimally close. The form of a rotation preserves the length of the real unit vector $q=\{q_\tau\}$ and therefore the normalization of the probability distribution,
\be \label{AA3}
q^2=\sum_\tau q_\tau^2=\sum_\tau p_\tau=1.
\ee
It is easy to describe periodic changes of $\{p_\tau\}$ in terms of such rotations \cite{3A}.

In the limit $\epsilon\to 0$ we may cast the rotation of $q$ in eq. \eqref{A} into the form of a differential equation
\be\label{4A}
\partial_tq=Kq~,~K=(\partial_tR)R^{-1}~,~K^T=-K.
\ee
For any time evolution of $q$ we can compute the time evolution of the probability distribution $\{p_\tau\}=\{q^2_\tau\}$. While the evolution equation \eqref{4A} for $q$ is a linear differential equation, the corresponding differential equation for $\{p_\tau\}$ can be non-linear. We may use the standard formalism of quantum mechanics by defining $K=-iH/\hbar $. In our approach the value of Planck's constant $\hbar$ is purely an issue of units. (The classical limit of quantum mechanics is realized as usual if the action is large in units of $\hbar$.) Using inverse time units for $H$ we will set $\hbar=1$.  The evolution is then described by a familiar type of ``Schr\"odinger equation''
\ba\label{30A}
\partial_t q&=&-iHq.
\ea
For finite $B$ the Hamiltonian $H$ is a purely imaginary antisymmetric $2^B\times2^B$ matrix and therefore hermitean, $H^\dagger=H$. At this point we have already the formal structure of the unitary time evolution of a wave function in quantum physics. We will see for detailed examples how such classical ensembles describe quantum systems. So far the wave function is real. We will explain in our specific examples how the complex structure characteristic for  quantum physics arises. We will concentrate here on the description of quantum systems for fermions, while bosonic objects can be found as composites involving even numbers of fermions. 

Starting from these rather simple and very general considerations we will see how the whole formalism for a quantum field theory of multi-fermion systems arises from a classical statistical ensemble. For the time evolution of the system we exploit the isomorphism between wave functions $\{q_\tau\}$ and elements of a Grassmann algebra. We construct a map between a family of wave functions $\{q_\tau(t)\}$ for every time $t$ on one side, and a Grassmann functional integral on the other side. This allows us to use all tools of quantum field theory for the description of a unitary time evolution of classical probabilities \eqref{A}. In particular, we formulate the time evolution for an Ising type classical statistical model in two dimensions which is equivalent to a two-dimensional Lorentz-invariant quantum field theory fermions. 

This paper is organized as follows. In sect. \ref{Grassmannrepresentation} we introduce the Grassmann representation of the wave function $\{q_\tau\}$. Up to irrelevant signs this also constitutes a Grassmann representation of the ``classical'' probability distribution $\{p_\tau\}$. In sect. \ref{fermions} we briefly recapitulate the connection between the Grassmann representation of the wave function and the quantum wave function of a multi-fermion system in Fock space. Sect. \ref{ConjugateGrassmannvariables} is devoted to conjugate Grassmann variables. We show how arbitrary Grassmann operators can be represented as multiplicative operators constructed as linear combinations of powers of the Grassmann variables $\psi$ and their conjugates $\hat{\psi}$. In sect. \ref{Symmetries} we discuss symmetry transformations of the probability distribution and wave functions.

The second part of this paper establishes the map between Grassmann functional integrals and unitary time evolutions of classical statistical ensembles. In sect. \ref{quantumwavefunction} we construct the wave function $\{q_\tau(t)\}$ for every time $t$ by a partial integration of the degrees of freedom of the Grassmann functional integral. We first use a very simple example corresponding to two-state quantum mechanics. In sect. \ref{cptsqm} we indicate the classical probability distribution and its time evolution which corresponds to the functional integral and  quantum wave function of sect. \ref{quantumwavefunction}. The opposite direction is followed in sect. \ref{conjugategrassmann}, where we start from the time evolution of classical probabilities and construct the associated Grassmann functional integral. We generalize our example and discuss the constraints that the Grassmann functional integral has to obey in order to permit a one to one correspondence with a unitary evolution of the wave function $\{q_\tau\}$. 

While the first example in sects. \ref{quantumwavefunction}-\ref{conjugategrassmann} employs a real Grassmann algebra, we present a simple example for a complex structure and the resulting complex Grassmann algebra in sect. \ref{Twocomponentspinor}. Sect. \ref{A16} addresses the issue of the boundary terms in the functional integral which specify the specific quantum state. If one of these terms is formulated as an initial condition for the past, while the other concerns the future, we need to relate the future to the past. This can be done using translations and reflection in time. Finally, in sect. \ref{qft} we discuss a two-dimensional quantum field theory for free fermions within our setting. We derive the time evolution of the classical probability distribution $\{p_\tau(t)\}$ from a Lorentz-invariant Grassmann functional integral. Inversely, for a suitable time evolution of the classical probability distribution for an Ising type system we construct the quantum wave function for a multi-fermion system and its time evolution. This can then be mapped to a functional integral, from which the observables in a quantum field theory can be computed in the usual way. This system admits a complex structure which can be associated to the notion of antiparticles.  We present our conclusions in sect. \ref{conclusions}.

\section{Grassmann representation of classical probabilities}
\label{Grassmannrepresentation}
We want to show that classical statistical ensembles with probability distribution $\{p_\tau\}$ following a ``unitary time evolution'' \eqref{4A} can describe quantum systems for fermions. In particular, we want to recover the standard formalism with creation and annihilation operators. For this purpose we first map in this section the classical wave function $\{q_\tau\}$ to an element of a Grassmann algebra, and connect the two-level observables $N_\alpha$ to operators acting within the Grassmann algebra. The connection between the Grassmann algebra and fermionic quantum systems is standard and will be briefly recapitulated in the next section. In sect. \ref{qft} we will construct a classical statistical ensemble describing a quantum field theory for fermions. From this we can easily obtain quantum field theories for bosons as suitable composite fields for fermion bilinears or more complicated bosonic composite fields. This may include gravity and gauge interactions as in spinor gravity \cite{HC}.

The map from the classical wave function to an element of a Grassmann algebra has been constructed in detail in ref. \cite{3A}. We briefly recapitulate in this section the representation of states, probabilities, observables and evolution law in terms of a Grassmann algebra. The two-level observables $N_\alpha$ can be associated with bits taking the values one and zero. 
To each bit we associate a Grassmann variable $\psi_\alpha,\alpha=1\dots B$. Grassmann variables anticommute and obey the standard differentiation and integration rules
\ba\label{22}
\psi_\alpha\psi_\beta+\psi_\beta\psi_\alpha&=&0,\\
\frac{\partial}{\partial\psi_\beta}\psi_\beta \psi_{\alpha_1}\dots\psi_{\alpha n}&=&
\int d\psi_\beta\psi_\beta\psi_{\alpha_1}\dots \psi_{\alpha_n}=
\psi_{\alpha_1}\dots\psi_{\alpha_n}.\nn
\ea
To every state $\tau=[n_\beta]$ we associate an element of the Grassmann algebra, $\tau\to g_\tau$, which is a product of factors $\psi_\beta$ according to the following rule: if within the sequence $[n_\beta]$ one has for a given $\alpha$ the value $n_\alpha=0$, we take a factor $\psi_\alpha$ in $g_\tau$, while $n_\alpha=1$ corresponds to a factor $1$. Thus an ordered chain $(n_1,n_2,\dots,n_B)$ corresponds to a product of factors $\psi_\alpha$ for each $n_\alpha=0$ in the chain. 
The Grassmann elements $g_\tau$ form a complete basis of the real Grassmann algebra $\cal{G}$.

To every wave function $\{q_\tau \}=\{s_\tau \sqrt{p_\tau}\}$ we associate an element of the Grassmann algebra $g\in {\cal G}$, 
\be\label{24}
g=\sum_\tau q_\tau g_\tau~.
\ee
We also define the conjugate element $\tilde g$ of $g$ by
\be\label{25}
\tilde g=\sum_\tau q_\tau \tilde g_\tau
\ee
where $\tilde g_\tau$ obtains from  $\tau=[n_\beta]$ by taking a factor $\psi_\alpha$ for every $n_\alpha=1$, and a factor $1$ for $n_\alpha=0$, with an appropriate sign such that $\tilde g_\tau g_\tau=\psi_1\psi_2\dots \psi_B$. More formally, we may define $\tilde g_\tau$ by the property
\be\label{26}
\int {\cal D}\psi\tilde g_\tau g_\rho=\delta_{\tau\rho},
\ee
where the integral over all Grassmann variables reads
\be\label{27}
\int {\cal D}\psi=\int d\psi_B\dots \int d\psi_2\int d\psi_1.
\ee
The elements $g$ and $\tilde g$ associated to the probability distribution $\{p_\tau\}$ are normalized according to
\be\label{28}
\int {\cal D}\psi~ \tilde g g=\int {\cal D}\psi\sum_{\tau',\tau} q_\tau\tilde g_\tau q_{\tau'} g_{\tau'}=\sum_\tau p_\tau=1.
\ee

We can realize a classical observable $A$ as an operator ${\cal A}$ acting on $g$ according to 
\be\label{29}
{\cal A} g_\tau=A_\tau g_\tau~,~
\int {\cal D}\psi\tilde g{\cal A} g=
\sum_\tau p_\tau A_\tau=\kl A\kr.
\ee
In particular, the occupation number operator ${\cal N}_\alpha$ associated to $N_\alpha$ reads
\be\label{30}
{\cal N}_\alpha=\partial_\alpha\psi_\alpha=\frac{\partial}{\partial\psi_\alpha}
\psi_\alpha.
\ee
Different occupation numbers commute, ${\cal N}_\alpha{\cal N}_\beta={\cal N}_\beta{\cal N}_\alpha$.

The element $g=\sum_\tau q_\tau g_\tau$ plays the role of a Grassmann-valued wave function, similar to a vector which obtains as a sum over basis elements $g_\tau$ with coefficients $q_\tau$. The time evolution of $\{q_\tau(t)\}$,
\be \label{13A}
q_\tau(t)=\sum_{\rho}R_{\tau\rho}(t, t_0)q_\rho(t_0)
\ee
is transfered to the time evolution of $g(t)$,
\be\label{42a}
g(t)={\cal U}(t,t_0)g(t_0)~,~\tilde g(t)=\tilde{{\cal U}}^T(t,t_0)\tilde g(t_0).
\ee
Here the Grassmann operator ${\cal U}$ plays a role similar to the unitary evolution operator in quantum mechanics, with $\tcu$ the inverse of ${\cal U}$,
\be\label{50}
\tcu (t,t_0) {\cal U}(t,t_0)= 1_{\cal G}.
\ee
(For more details see ref. \cite{3A}.) In the Grassmann formulation the Hamilton operator ${\cal H}$ is defined as
\be\label{113A}
{\cal H}(t)=i\partial_t{\cal U}(t,t_0)\tcu(t,t_0),
\ee
and the time evolution \eqref{42a} can be written as a Schr\"odinger equation
\be\label{113B}
\partial_tg(t)=-i{\cal H}(t)g(t).
\ee
For a real Grassmann algebra the operator ${\cal K}=-i{\cal H}$ is real such that the appearance of $i$ in eq. \eqref{113B} serves only for the formal analogy to quantum mechanics. 

If a suitable complex structure exists we can map the elements of a real Grassmann algebra to elements of a complex Grassmann algebra with $2^{B-1}$ basis elements. For example, it may be built from $F=B-1$ Grassmann variables. Such mappings will be discussed in our explicit examples. They correspond to maps from a real vector $\{q_\tau \}$ to  a complex vector $\{c_\tau\}$ with half the number of components. The probabilities read now $p_\tau=|c_\tau|^2$. Replacing in eq. \eqref{25} $q_\tau\to c_\tau^*$ the normalization \eqref{28} remains valid for the complex Grassmann algebra as well. The time evolution \eqref{13A} or \eqref{42a} will then be mapped to the unitary time evolution of a complex wave function $\{c_\tau \}$, with $R_{\tau\rho}$ replaced by a unitary matrix $U_{\tau\rho}$. Within the description by a complex Grassmann algebra ${\cal H}$ needs no longer to be purely imaginary.

\section{Fermions} 
\label{fermions}
In this section we recall how each element of the Grassmann algebra can be associated with the quantum wave function for a multi-fermion system. This material is rather standard and displayed here only for completeness and for setting the notation for later sections. We will work here with a complex Grassmann algebra where the elements can be labeled by the components $c_\tau$ of a complex vector. (The case of a real wave function and real Grassmann algebra is recovered by using real $c_\tau=q_\tau$.) The multi-fermion system can be described in the occupation number basis. For only one species of fermions the Hilbert space can be built from two states, the vacuum $|0\rangle$ (empty state) and the one fermion state $|1\rangle$ (occupied state). The occupied state obtains by applying the creation operator $a^\dagger$ to the vacuum
\be\label{F1}
|1\rangle=a^\dagger|0\rangle,
\ee
where the annihilation operator $a$ acting on the occupied state yields the vacuum
\be\label{F2}
a|1\rangle=|0\rangle.
\ee
Using further the relations
\be\label{F3}
a|0\rangle=a^\dagger|1\rangle=0
\ee
implies $a^2=(a^\dagger)^2=0$. 
The occupation number operator ${\cal N}=a^\dagger a$ acts as 
\be\label{F4}
a^\dagger a| 1\rangle= | 1\rangle,  \       a^\dagger a |0\rangle=0.
\ee
Thus $| 1\rangle$ and $|0\rangle$ are eigenstates of ${\cal N}$ with eigenvalues one and zero. Furthermore, with 
\be\label{F5}
aa^\dagger|1\rangle=0,  \ aa^\dagger|0\rangle=|0\rangle,
\ee
we infer the commutation relation
\be\label{F6} 
\left\{a^\dagger,a\right\}=1.
\ee
The wave function for a pure quantum state for one species of fermions
 can be written as
\be\label{F7} 
\varphi=c_1|1\rangle+c_0|0\rangle,  \  |c_1|^2+|c_0|^2=1.
\ee
Expressed in terms of a two-component complex vector $(c_1,c_0)$ one has the explicit representation 
\be\label{F8}
a=\left(\begin{array}{ll}
0,&0\\ 1,&0
\end{array}\right) , \  
a^\dagger=\left(\begin{array}{ll}
0,&1\\ 0, &0
\end{array}\right).    
\ee

The operators and states for one fermion species can be identified with the Grassmann operators and elements of the Grassmann algebra built from a single Grassmann variable $\psi$. We associate
\be\label{F9}
a=\psi, \ a^\dagger=\frac{\partial}{\partial\psi}, \ {\cal N}=\frac{\partial}{\partial\psi} \ \psi,
\ee
where the Grassmann operators act on states represented by elements of the Grassmann algebra. The basis elements are associated as 
\be\label{F10}
|0\rangle \widehat{=} \psi~ ,~ |1\rangle \widehat{=}1
\ee
such that
\be\label{F11}
 \varphi=c_1|1\rangle +c_0|0\rangle\, \widehat{=}\, g=c_1\psi+c_0.
\ee
In other words, $g$ is specified by the two complex numbers $\{c_\tau\}=\{c_1,c_0\}$ which characterize the quantum wave function \eqref{F7}.

The description of the quantum wave function for fermions by an element of the Grassmann algebra is easily extended to several fermion species that we label by $\alpha=1\dots F$. We now use a Grassmann algebra based on $F$ Grassmann variables $\psi_\alpha$. 
The basis states of the Hilbert space (Fock space) for the multi-fermion system obtain by applying various powers of creation operators $a^\dagger_\alpha$ for different fermions species to the vacuum. For the example of two species $(F=2)$ we have four basis states, the vacuum $|0,0\kr$, the two states $|1,0\kr$ and $|0,1\kr$ with one fermion, either of type $1$ or type $2$, and the totally occupied state $|1,1\kr$. They obey 
\ba\label{F12}
|1,0\kr&=&a^\dagger_1|0,0\kr~,~|0,1\kr=a^\dagger_2|0,0\kr,\nn\\
|1,1\kr&=&-a^\dagger_1 a^\dagger_2|0,0\kr=a^\dagger_2a^\dagger_1|0,0\kr\nn\\
&=&-a^\dagger_1|0,1\kr=a^\dagger_2|1,0\kr,
\ea
and
\ba\label{F13}
|1,0\kr&=&a_2|1,1\kr~,~|0,1\kr=-a_1|1,1\kr\nn\\
|0,0\kr&=&a_1a_2|1,1\kr=-a_2a_1|1,1\kr\nn\\
&=&a_2|0,1\kr=a_1|1,0\kr.
\ea
Our sign conventions for the basis states are compatible with the anti-commutation relations
\be\label{F14}
\{a_\alpha,a_\beta\}=0~,~\{a^\dagger_\alpha,a^\dagger_\beta\}=0~,~
\{a^\dagger_\alpha,a_\beta\}=\delta_{\alpha\beta}.
\ee

A pure state quantum wave function reads
\ba\label{F15}
\varphi=\sum_\tau c_\tau\varphi_\tau,
\ea
with
\be\label{F16}
\varphi_1=|1,1\kr~,~\varphi_2=|0,1\kr~,~\varphi_3=|1,0\kr~,~\varphi_4=|0,0\kr.
\ee
The map to the Grassmann algebra employs the identification
\be\label{F17}
a^\dagger_\alpha=\frac{\partial}{\partial\psi_\alpha}~,~a_\alpha=\psi_\alpha~,~
{\cal N}_\alpha=\frac{\partial}{\partial\psi_\alpha}\psi_\alpha,
\ee
and the association 
\ba\label{F18}
|0,0\kr~\widehat{=}~\psi_1\psi_2=g_4~&,&~|1,0\kr~\widehat{=}~\psi_2=g_3,\nn\\
|0,1\kr~\widehat{=}~-\psi_1=g_2~&,&~|1,1\kr~\widehat{=}~1=g_1,
\ea
with
\be\label{F19}
\varphi~\widehat{=}~g=\sum_\tau c_\tau g_\tau.
\ee
The relations \eqref{F14}, \eqref{F15}, \eqref{F17}, \eqref{F19} hold for an arbitrary number of fermion species $F$. Our conventions are such that states which obtain from the vacuum by applying an ordered sequence of creation operators $a^\dagger_\alpha$, with larger $\alpha$ to the left, have a plus sign. This concludes the one to one correspondence between multi-fermion wave functions and elements of a complex Grassmann algebra.

\section{Conjugate Grassmann variables}
\label{ConjugateGrassmannvariables}

The representation of Grassmann operators in terms of $\psi$ and $\partial/\partial\psi$ is somewhat cumbersome, since these two objects do not anticommute. We therefore discuss in this section another representation of the classical probability distribution, wave function and observables where $\partial/\partial\psi$ is effectively replaced by a new Grassmann variable $\hat\psi $, obeying
\be\label{V1}
\{\hat\psi _\alpha,\hat\psi _\beta\}=
\{\hat\psi _\alpha,\psi_\beta\}=
\{\psi_\alpha,\psi_\beta\}=0.
\ee

\bigskip\noindent
{\bf 1. Conjugate variables}

\medskip\noindent
This goal will be achieved by a type of Fourier transformation of $\tilde g$,
\be\label{V2}
\tilde g(\psi)=\int{\cal D}\hat\psi  
\exp(\sum_\beta\hat\psi _\beta\psi_\beta)
\hat g (\hat\psi ),
\ee
with
\be\label{V2a}
\int{\cal D}\hat\psi =\int d\hat\psi _1\int d\hat\psi _{2\dots}\int d\hat\psi _B.
\ee
We will systematically replace $\tilde g$ in sect. \ref{Grassmannrepresentation} by the associated element $\hat g (\hat\psi )$. 

The map $\hat{g}\to\tilde{g}$ in eq. \eqref{V2} is invertible and we next construct the inverse map $\tilde g \to \hat g $. Basic identities needed for this purpose can be found, for example, in ref. \cite{Zin}. For general Grassmann variables obeying eq. \eqref{V1} we note the identities (no sum over $\beta$ here)
\ba\label{V3}
&&\exp (\hat\psi _\beta\psi_\beta)=1+\hat\psi _\beta\psi_\beta,\nn\\
&&\int d\psi_\beta d\hat\psi _\beta\exp (\hat\psi _\beta\psi_\beta)=1,\nn\\
&&\frac{\partial}{\partial\psi_\beta}\exp\left(\hat\psi _\beta\psi_\beta\right)=-
\hat\psi _\beta\exp (\hat\psi _\beta\psi_\beta).
\ea
For a given $\alpha,\beta$ one has a $\delta$-function like object
\ba\label{V3A}
&&\delta(\hat\psi _\alpha-\hat\psi _\beta)=\hat\psi _\alpha-\hat\psi _\beta=-\delta(\hat\psi _\beta-\hat\psi _\alpha),\nn\\
&&\int d\hat\psi _\alpha\delta(\hat\psi _\alpha-\hat\psi _\beta)h(\hat\psi _\alpha)=h(\hat\psi _\beta),
\ea
where $h$ is an arbitrary element of the Grassmann algebra which may also involve Grassmann variables different from $\hat\psi _\alpha$. We can write
\ba\label{V3B}
\delta(\hat\psi _\alpha-\hat\psi _\beta)
&=&\int d \psi_\beta\exp \big\{\psi_\beta(\hat\psi _\alpha-\hat\psi _\beta)\big\}\\
&=&-\int d\psi_\beta\exp (-\hat\psi _\beta\psi_\beta)\exp(\hat\psi _\alpha\psi_\beta),\nn
\ea
and 
\ba\label{102A}
\delta(\psi_\alpha-\psi_\beta)&=&\int d\hat\psi _\beta \exp 
\big\{\hat\psi _\beta(\psi_\alpha-\psi_\beta)\big\}\nn\\
&=&-\delta(\psi_\beta-\psi_\alpha).
\ea
These identities are easily generalized
\ba\label{V3C}
&&\int \cD \psi\cD \hat\psi \exp (\sum_\beta\hat\psi _\beta\psi_\beta)=1\nn\\
&&\frac{\partial}{\partial\psi_\alpha}\exp (\sum_\beta\hat\psi _\beta\psi_\beta)
=-\hat\psi _\alpha\exp (\sum_\beta\hat\psi _\beta\psi_\beta),\nn\\
&&\delta (\hat\psi '-\hat\psi )=(-1)^B\int\cD\psi\exp 
\big \{\sum_\beta (\hat\psi '_\beta-\hat\psi _\beta)\psi_\beta\big \},\nn\\
&&\int \cD\hat\psi '\delta(\hat\psi '-\hat\psi )h(\hat\psi ')=h(\hat\psi ).
\ea

The identities \eqref{V3C} can be used in order to verify that the inverse transformation $\tilde g\to \hat g $ obeys
\be\label{V3D}
\hat g (\hat\psi )=\int \cD\psi\exp 
\left(-\sum_\beta\hat\psi _\beta\psi_\beta\right)\tilde g(\psi).
\ee
Since eqs. \eqref{24}, \eqref{25} (and similar for a complex Grassmann algebra) define a map $g\to\tilde g$, we can combine this with eq. \eqref{V3D} for the definition of a map $g\to\hat g $.
In fact, the element $\hat g (\hat\psi )$ can be easily obtained from $g(\psi)$ by the following steps: (i) each factor $\psi_\alpha$ in $g$ is replaced by $\hat\psi _\alpha$ in $\hat g $, (ii) the factors $\hat\psi _\alpha$ are totally reordered in $\hat g $, (iii) for a complex Grassmann algebra one takes the complex conjugate of the coefficients $c_\tau$. For example, the element associated to $g=c\psi_1\psi_2$ is $\hat g =c^*\hat\psi _2\hat\psi _1$. The normalization of the conjugate basis elements $\hat g_\tau$ obeys
\be\label{V3E}
\int \cD\psi\int \cD\hat\psi \exp (\sum_\beta\hat\psi _\beta\psi_\beta)
\hat g _\rho(\hat\psi ) g_\tau (\psi)=\delta_{\rho\tau}.
\ee
We also note the useful identities
\ba\label{82A}
\sum_\tau g_\tau (\psi)\hat g _\tau(\hat\psi )=\exp 
(\sum_\beta \psi_\beta\hat\psi _\beta),\nn\\
\sum_\tau\hat g _\tau(\hat\psi )g_\tau(\psi)=\exp 
(\sum_\beta\hat\psi _\beta\psi_\beta).
\ea
In terms of the conjugate basis elements we can write
\be\label{49A}
g=\sum_\tau c_\tau g_\tau~,~\hat g=\sum_\tau c^*_\tau\hat g_\tau,
\ee
with $c_\tau=c^*_\tau=q_\tau$ for a real Grassmann algebra.

\bigskip\noindent
{\bf 2. Conjugate Grassmann operators}

\medskip\noindent
We next define for an arbitrary Grassmann operator ${\cal F}$ the normal ordered form $\cF_N$. It consists of sums of products of factors $a+b\psi_\beta+c\partial/\partial\psi_\beta+d(\partial/\partial\psi_\beta)\psi_\beta$, i.e. where $\partial/\partial\psi_\beta$ always precedes a possible factor $\psi_\beta$. We will replace $\cF_N(\partial/\partial\psi_\beta,\psi_\beta)$ by $F_N(\hat\psi _\beta,\psi_\beta)$
according to 
\ba\label{V7}
&&\int \cD\psi\tilde g(\psi)
{\cal F}_N\left(\frac{\partial}{\partial\psi},\psi\right)g(\psi)\nn\\
&&=\int \cD\psi \cD\hat\psi \exp (\sum_\beta\hat\psi _\beta\psi_\beta)
\hat g (\hat\psi )F_N(\hat\psi ,\psi)g(\psi).
\ea
Here $F_N(\hat\psi,\psi)$ obtains from $\cF_N\left(\frac{\partial}{\partial\psi},\psi\right)$ by replacing $\partial/\partial\psi\to\hat\psi$. The proof of the identity \eqref{V7} can be sketched as follows. By partial integration we find for an arbitrary element $f$ of the Grassmann algebra
\ba\label{V4a}
&&\int \cD\psi\tilde g(\psi)\frac{\partial}{\partial\psi_\alpha}
f(\psi)\nn\\
&&=\int\cD\psi\cD\hat\psi \exp (\sum_\beta\hat\psi _\beta\psi_\beta)
\hat g (\hat\psi )\hat\psi _\alpha f(\psi).
\ea
The identity \eqref{V4a} also holds if $f$ takes the form $f(\psi)=$ 
${\cal F}\left(\frac{\partial}{\partial \psi_\beta},\psi_\beta\right)g(\psi)$, with $\cF$ an arbitrary Grassmann operator expressed in terms of $\psi_\beta$ and derivatives $\partial/\partial\psi_\beta$. Furthermore, we generalize eq. \eqref{V4a} to
\ba\label{V5}
&&\int \cD\psi \cD\hat\psi \exp (\sum_\beta\hat\psi _\beta\psi_\beta)
h(\psi,\hat\psi )\frac{\partial}{\partial\psi_\alpha}f(\psi)\nn\\
=&&\int \cD\psi\cD\hat\psi \exp (\sum_\beta\hat\psi _\beta\psi_\beta)h(\psi,\hat\psi )\hat\psi _\alpha f(\psi)
\ea
for all elements $h$ obeying $\partial h/\partial\psi_\alpha=0$. Using the anticommutation relation
\be\label{V6}
\left\{\frac{\partial}{\partial\psi_\alpha}, \psi_\beta \right\}=\delta_{\alpha\beta}
\ee
we can bring any arbitrary Grassmann operator ${\cal F}$ into its normalized form ${\cal F_N}$. From the identities \eqref{V4a}, \eqref{V5} for these ordered operators we finally obtain the relation \eqref{V7}. 

In particular, the number operators \eqref{30} read in this representation
\be\label{V8}
{N}_\alpha=\hat\psi _\alpha\psi_\alpha,
\ee
while polynomials of different ${\cal N}_{\alpha_1}, {\cal N}_{\alpha_2}$ etc. are  represented by corresponding polynomials of $\hat\psi_{\alpha_1}\psi_{\alpha_1}$, $\hat\psi _{\alpha_2}\psi_{\alpha_2}$ etc. (Note, however, that ${\cal N}^2_\alpha$ is not represented by $(\hat\psi _\alpha\psi_\alpha)^2=0$, since the normal ordered form reads $({\cal N}^2_\alpha)_N={\cal N}_\alpha$.) The annihilation and creation operators \eqref{F9} are now represented by $\psi$ and $\hat\psi $. 

In conclusion, we have arrived at a representation of the classical probability distribution and observables where the probability distribution $\{p_\tau\}$ is represented by the conjugate pair of Grassmann elements $g(\psi)$ and $\hat g (\hat\psi )$, and the observables are polynomials in $\psi$ and $\hat\psi $. The elements $g,\hat g $ are normalized according to 
\ba\label{V9}
&&\int D\psi\hat g (\hat\psi )g(\psi)=1,\nn\\
&&\int D\psi=\int \cD \psi\int \cD\hat\psi \exp(\sum_\beta\hat\psi _\beta\psi_\beta).
\ea
They encode the information about the classical ensemble $\{p_\tau\}$ according to 
\be\label{V10}
c_\tau=\int D\psi\hat g _\tau (\hat\psi ) g(\psi)~,~p_\tau=|c_\tau|^2.
\ee
For a time dependent probability distribution $\{p_\tau(t)\}$ the coefficients $c_\tau(t)$, and therefore $g(t)=\sum_\tau c_\tau(t)g_\tau$, depend on $t$. 

A classical observable $A$ is represented by an element of the Grassmann algebra ${A}(\hat\psi,\psi)$ that does not contain derivatives anymore. For example, we may investigate the (equal time) correlation function of the classical statistical ensemble,
\be \label{50A}
\bar C^{(m)}_{\alpha_1\dots\alpha_m}(t)=\sum_{\tau}p_\tau N_{\alpha_1,\tau}\dots N_{\alpha_m,\tau},
\ee
with $N_{\alpha_k,\tau}=n_{\alpha_k}$ the value of the observable $N_{\alpha_k}$ for the state $\tau$. In terms of the Grassmann algebra this correlation function is given by
\be\label{V11}
\bar C^{(m)}_{\alpha_1\dots\alpha_m}(t)=\int D\psi\hat g (\hat\psi ;t)
{N}_{\alpha_1}\dots {N}_{\alpha_m}g(\psi;t),
\ee
with ${N}_\alpha=\hat\psi _\alpha\psi_\alpha$, provided that all $\alpha_j$ are different from each other.

\section{Symmetries}
\label{Symmetries}
At a given time $t$ a classical statistical ensemble is characterized by a probability distribution $\{p_\tau\}$ which corresponds to a map $\tau\to p_\tau$. Consider an invertible transformation between the states, $\tau'=S(\tau)~,~\tau=S^{-1}(\tau')$. This induces a transformation in the space of statistical ensembles, $\{p_\tau\}\to \{p'_\tau\}=S\big(\{p_\tau\}\big)$, given by $p'_{\tau'}=p_{\tau(\tau')}=p_{S^{-1}(\tau')}$. The transformation $S$ is a symmetry if the probability distribution remains uncharged, $p'_\tau=p_\tau$. In this case the transformed observable $A'=S(A)$, which takes for every state $\tau$ the value $A'_\tau=A_{\tau'(\tau)}=A_{S(\tau)}$, has the same expectation value as $A$,
\ba\label{S1a}
\kl A'\kr&=&\sum_\tau A'_\tau p_\tau=\sum_\tau A_{\tau'(\tau)}p_\tau\nn\\
&=&\sum_{\tau'}A_{\tau'}p_{\tau(\tau')}=
\sum_{\tau'}A_{\tau'}p'_{\tau'}=\kl A\kr.
\ea
In the presence of an evolution law the notion of symmetry can be extended to relate probability distributions at different times,
\be\label{S2a}
\big\{p'_\tau(t_1)\big\}=\big\{p_\tau(t_0)\big\}
\Rightarrow \kl A'(t_1)\kr=\kl A(t_0)\kr.
\ee

We will be interested in symmetries where $S(\tau)$ is realized by simple transformations of the occupation numbers $n_\alpha$. The transformation of ``charge conjugation'' maps $C(n_\alpha)=\bar n_\alpha=1-n_\alpha$, inducing $C(\tau)=\bar\tau$ where $\bar\tau$ obtains from $\tau$ by flipping all bits. A time reversal combines this transformation with a reflection in time around some point $t_0$, $p_{\bar\tau}(t_0+t)=p_\tau(t_0-t)$. Symmetry under charge conjugation $C$ or time reversal $T$ implies $\kl C(A)(t)\kr=\kl \bar A(t)\kr=\kl A(t)\kr$ or $\kl T(A)(t_0+t)\kr=\kl \bar A(t_0+t)\kr=\kl A(t_0-t)\kr$ with $\bar A_\tau=A_{\bar\tau }$.

In the Grassmann formalism the map $\tau\to S(\tau)$ is reflected by a map of the basis elements $g_\tau\to g'_\tau=\pm g_{\tau'}=\pm g_{S(\tau)}=s_{\tau'} g_{\tau'}$. The induced map for an arbitrary Grassmann element $g\to g'=\sum_\tau q_\tau g'_\tau$ can equivalently be expressed by a transformation of the wave function $\{q_\tau\}$ at fixed $g_\tau$, 
\be\label{S3a}
S(g)=g'=\sum_\tau q_\tau g'_\tau=\sum_\tau q'_\tau g_\tau,
\ee
with  $q'_{\tau'}=s_{\tau'}q_{\tau(\tau')}$. We can represent the transformation $S$ by a matrix multiplication
\be\label{S4a}
q'_\tau=\sum_\rho S_{\tau\rho}q_\rho
\ee
where $S$ has only elements $1,-1$ and $0$, with precisely one value $1$ or $-1$ in each row and column, $S_{\tau\rho}S_{\tau\rho'}=|S_{\tau\rho}|\delta_{\rho\rho'}$. With $(S_{\tau\rho})^2=|S_{\tau\rho}|$ this guarantees
\be\label{S5a}
p'_\tau=\sum_\rho|S_{\tau\rho}|p_\rho.
\ee
For a fixed wave function the basis elements transform as
\be\label{S6a}
g'_\tau=\sum_\rho(S^T)_{\tau\rho}g_\rho=\sum_\rho g_\rho S_{\rho\tau}.
\ee
A symmetry is realized if $\{q_\tau\}$ is an eigenvector of $S$ with eigenvalue one. This has to hold for a suitable choice of signs $s_{\tau'}$. 

The transformations of $\tau$ which are induced by transformations of the occupation numbers $n_\alpha$ may be expressed as transformations of the Grassmann variables $\psi_\alpha$. Consider first a transformation of the type
\be\label{S7a}
S(\psi_\alpha)=\psi'_\alpha=\sum_\beta W_{\alpha\beta}\psi_\beta~,~
W^T W=1.
\ee
This defines a transformation $S(g_\tau)=g'_\tau$ of the basis elements by replacing in $g_\tau$ each factor $\psi_\alpha$ by $\psi'_\alpha$. The resulting matrix $S_{\tau\rho}$ in eq. \eqref{S6a} is orthogonal, which guarantees that the normalization of the probability distribution is preserved. For the particular case where all matrix elements $W_{\alpha\beta}$ equal $1,-1$ or $0$, with only one element $1$ or $-1$ in each row or column, this property is transfered to the matrix $S_{\tau\rho}$. We recover the map of classical states $\tau\to \tau'$, as induced by $n'_\alpha=\sum_\beta|W_{\alpha\beta}|n_\beta$. 

We emphasize, however, that {\em every} transformation \eqref{S7a} with orthogonal $W$ induces a transformation $S(g_\tau)$ and corresponding transformations $S\big(\{q_\tau\}\big)$ and $S\big(\{p_\tau\}\big)$. In particular, if $S\big(\{p_\tau\}\big)=\{p_\tau\}$ we find symmetry transformations of the probability distribution that go beyond the ones realized by a mapping of states $\tau\to\tau'$. For example, the system may realize a symmetry originating from continuous rotations of the vector $\{\psi_\alpha\}$. No corresponding simple construction as a ``rotation among states'' is available on the level of the occupation numbers $n_\alpha$. The Grassmann formalism can be a powerful tool for realizing simple symmetries of a classical statistical ensemble. More generally, the group of all orthogonal matrices $S_{\tau\rho}$ in eq. \eqref{S4a}, which leave a given wave function $\{q_\tau\}$ invariant, may be considered as symmetries. On the level of $\{p_\tau\}$ this may correspond to non-linear transformations. 

Furthermore, we may investigate transformations that map Grassmann elements to conjugate ones. An example is the charge conjugation which transforms
\be\label{S8a}
\cC \psi_\alpha=\tilde\psi_\alpha,
\ee
where $\tilde\psi_\alpha$ is the conjugate Grassmann element of $\psi_\alpha$, with 
\be\label{S9a}
\int {\cal D}\psi\tilde\psi_\alpha\psi_\beta=\delta_{\alpha\beta}.
\ee
The basis elements are mapped by charge conjugation as
\be\label{S10a}
\cC g_\tau=g^c_\tau=\pm \tilde g_\tau~,~C(g)=\sum_\tau q_\tau g^c_\tau=\sum_\tau
q^c g_\tau.
\ee
A charge conjugation invariant state obeys $q^c_\tau=q_\tau$ for an appropriate choice of the sign of $g^c_\tau$. It realizes a symmetry of the classical probability distribution $\{p^c_\tau\}=\big\{(q^c_\tau)^2\big\}=\{q^2_\tau\}=\{p_\tau\}$. For a classical observable $A$ constructed from the two-level observables (e.g. equal time classical correlation functions) the charge conjugate observable $A^c=C(A)$ obtains by replacing each factor $N_\alpha$ by $1-N_\alpha$. The associated Grassmann operator ${\cal A}^c$ obtains from ${\cal A}$ by replacing $\psi_\alpha\leftrightarrow \partial/\partial\psi_\alpha$. For a charge conjugation invariant state one finds 
$\kl A^c\kr=\kl A\kr$. 

The concept of symmetries is extended to the formalism with conjugate Grassmann variables in a straightforward way. Transformations of the type \eqref{S7a} can be associated to corresponding transformations of the wave function \eqref{S4a} and therefore to the transformation of the conjugate Grassmann elements
\be\label{68A}
\hat g'=\sum_\tau q'_\tau\hat g_\tau=\sum_\tau q_\tau\hat g'_\tau.
\ee
In the second equation we transform the conjugate basis elements $\hat g_\tau$, keeping the wave function fixed. This general transformation rule for $\hat g_\tau$ can be extended to arbitrary transformations \eqref{S4a}, including, for example, a charge conjugation of the type \eqref{S8a}. The conjugate Grassmann variables $\hat \psi_\alpha$ are particular conjugate basis elements and their transformation properties are therefore fixed. Transformations of the type \eqref{S7a} transform basis elements containing only one factor $\psi_\alpha$ into each other
\ba\label{68B}
g^{(1)}=\sum_\alpha q^{(1)}_\alpha\psi_\alpha~,~g^{(1)}{'}=\sum_{\alpha,\beta}q^{(1)}_\alpha
W_{\alpha\beta}\psi_\beta=\sum_\beta q^{(1)}{'}_\beta\psi_\beta.\nn\\
\ea
With $\hat g^{(1)}=\sum_\alpha q^{(1)}_\alpha\hat\psi_\alpha$ we find the transformation
\be\label{68C}
S(\hat\psi_\alpha)=\hat\psi'_\alpha=\sum_\beta W_{\alpha\beta}\hat\psi_\beta.
\ee
As a result, the factor $\exp(\sum_\beta \hat\psi_\beta\psi_\beta)$ appearing in the transformations \eqref{V2}, \eqref{V3D} or the measure \eqref{V9} is invariant under such transformations.

\section{Quantum wave function from Grassmann functional integral}
\label{quantumwavefunction}
It is well-known that quantum systems of fermions can be described by Grassmann functional integrals. In the opposite direction, a Grassmann functional integral obeying certain conditions naturally leads to a wave function $\{q_\tau (t) \}$ obeying a unitarity time evolution, and the associated classical probability distribution $\{p_\tau (t) \}$. In this section we present a simple example how the unitary time evolution follows from a suitable Grassmann functional integral. In the following two sections we take the opposite direction, starting with a given time evolution of a probability density $\{p_\tau (t) \}$, and inferring the associated Grassmann functional integral.  We will present further examples in sects. \ref{Twocomponentspinor} and \ref{qft}.

\bigskip\noindent
{\bf 1. Action and functional integral}

For a formulation of a Grassmann functional integral we consider a discrete chain of different times $t_n$, with $t_{n+1}-t_n=\epsilon$ and $t_{in}\le t_n \le t_f$. In the end we take the limit $\epsilon\to 0$ for fixed $t_{in}$ and $t_f$. For each $t_n$  we employ a set of $B$ Grassmann variables $\psi_\alpha(t_n)$ and the associated $\hat{\psi}_\alpha(t_n)$. The action $S$ is an element of the real Grassmann algebra constructed from $\psi_\alpha(t_n)$ and $\hat{\psi}_\alpha(t_n)$. As  a simple example with $B=1$ we discuss the action
\be\label{101A}
S=\sum_{t'}L(t')
\ee
with
\ba\label{101B}
L(t')=\hat\psi (t')\big(\psi(t'+\epsilon)-\psi(t')\big)
+\epsilon\omega\big (\hat\psi (t')-\psi(t'+\epsilon)\big).\nn\\
\ea
The sum over $t'$ extends over all $t_n$ within the interval $t_{in}\le t_n \leq  t_f$, where for $t'=t_f$ we define $L(t_f)=-\hat\psi (t_f)\psi(t_f)$. Taking the continuum limit $\epsilon\to 0$ the action takes the form
\be\label{117XA}
S=\int dt\Big\{\hat\psi (t)\partial_t\psi(t)+\omega\big(\hat\psi (t)-\psi(t)\big)\Big\}.
\ee

We may define the hermitean conjugate of a Grassmann element $G[\psi,\hat\psi ]$ by exchanging $\psi\leftrightarrow\hat\psi $, a total transposition of the order of all Grassmann variables, and a complex conjugation of all coefficients, $G^{\dagger}[\psi, \hat\psi ]=\left(G^*[\hat\psi ,\psi] \right)^T$. With this definition we observe that $S$ is antihermitean, $S^\dagger=-S$. This corresponds to a hermitean Minkowski action
\be\label{117XB}
S_M=iS~,~e^{-S}=e^{iS_M},
\ee
as usual for path integrals in quantum mechanics. 

The partition function is defined as
\ba\label{M1A}
Z=\int \cD\psi(t')\cD\hat\psi (t')\hat g\big (\hat\psi (t_f)\big)\hat T
\{e^{-S[\psi,\hat\psi ]}\}g\big (\psi(t_{in})\big).\nn\\
\ea
Here we have introduced a time ordering $\hat T$ which puts factors $e^{-L(t_a)}$ to the left of factors $e^{-L(t_b)}$ if $t_a>t_b$. This specification is necessary for our example because $L(t')$ contains a term with an odd number of Grassmann variables such that factors $e^{-L(t_a)}$ and $e^{-L(t_b)}$ do not commute. If $S$ contains only terms with an even number of Grassmann variables there is no need for such ordering. The functional integration extends over all Grassmann variables $\psi(t')$, $\hat{\psi}(t')$, $t_{in}\le t' \le t_f$
\be \label{M1A1}
\int\cD\psi(t')\cD\hat\psi (t')=\prod_{t'}\int d \psi(t')d \hat\psi (t').
\ee
The factor $g\big (\psi(t_{in})\big)$ is a ``boundary term'' which will correspond to the initial condition for the wave function $q_\tau(t_{in})$. We will see that the functional integral defines a wave function $q_\tau(t)$ only if
$\hat g\big (\hat\psi (t_f)\big)$ is properly related to $g\big (\psi(t_{in})\big)$. This will restrict the functional integrals which describe a wave function $q_\tau(t)$. 

We can associate to $S$ the normalized action
\ba\label{M1}
S_N[\psi(t'),\hat\psi (t')]=\sum_{t'}L(t')+\ln Z=S+\ln Z,
\ea 
and define $G$ in terms of $S_N$
\be\label{M1B}
G[\psi,\hat\psi ]=\hat g\big (\hat\psi (t_f)\big)\hat T
\{e^{-\tilde S_N[\psi,\hat\psi ]}\}g\big (\psi(t_{in})\big).
\ee
This Grassmann element is normalized according to
\ba\label{104A}
\int \cD\psi(t')\cD\hat\psi (t')G
[\psi,\hat\psi ]=1.
\ea
We will see below that we can achieve $Z=1$ by a proper normalization of $g_{in}=g\big(\psi(t_{in})\big)$, such that no need for the shift \eqref{M1} arises. 

In order to have a well defined Grassmann functional integral we use a finite number of time steps  $t_{in}\leq t'\leq t_f$. Without loss of generality, we choose here $t_{in}=-t_f$. The functional integral will then be defined as the limit $\lim(t_f\to\infty)\lim(\epsilon\to 0)$, or we may consider $\epsilon\to 0$ for some finite $t_f$. The Grassmann element $G$ as well as $Z$ and other quantities depend on the boundary terms $g_{in}=g\big(\psi(t_{in})\big)$ and $\hat g_f=\hat g\big (\hat\psi (t_f)\big)$ and we will discuss the appropriate relation between $\hat g_f$ and $g_{in}$ below. The action \eqref{101A} is invariant under a time reflection, accompanied by a simultaneous exchange of $\psi$ and $\hat\psi $, a change of sign of $\psi,\hat\psi $, and a total reordering of all Grassmann variables
\be\label{M2a}
\big (S[-\hat\psi (-t'),-\psi(-t')]\big)^T=
S[\psi(t'),\hat\psi (t')].
\ee

\bigskip\noindent
{\bf 2. Wave function}

\medskip\noindent
The expectation value of observables at a given time $t$ involve only the probability distribution at time $t$. In other words, it involves only the information encoded in the Grassmann element $g(t)$. In contrast, the functional integral \eqref{104A} involves the Grassmann element $G$. This belongs to a much larger Grassmann algebra which is built from the variables $\psi(t')$ and $\hat\psi(t')$ for all $t'$. We therefore construct a map $G\to g(t)$ which focuses on the necessary information. 

We start with the functional integral expression for the occupation number $N(t)$
\ba\label{M2Aa}
\kl N(t)\kr=Z^{-1}\int \cD\psi(t')\cD\hat\psi (t')\hat g_f N(t)\nn\\
\times\hat T \{\exp\big (-S[\psi(t'),\hat\psi (t')]\big)\}g_{in},
\ea
with
\be\label{84A}
N(t)=\hat\psi(t)\psi(t).
\ee
We want to show that this equals the expectation value of the two-level observable $N(t)$ in terms of a wave function $q_\tau(t)$. For this purpose we decompose
\be\label{M3}
S=S_<+S_>-\hat\psi (t)\psi(t),
\ee
with
\ba\label{M4}
S_<&=&\sum_{t'<t}L(t'),\nn\\
S_>&=&\sum_{t'\geq t}L(t')+\hat\psi (t)\psi(t).
\ea
This decomposition is chosen such that $S_<$ depends only on $\psi(t'<t),\hat\psi (t'<t)$ and $\psi(t)$, while $S_>$ depends on $\psi(t'>t),\hat\psi (t'>t)$ and $\hat\psi (t)$. We will show that $q_\tau(t)$ or the associated Grassmann element $g(t)$ obtains by a suitable functional integral over Grassmann variables at $t'<t$, which involves $S_<$.

Using the decomposition \eqref{M4} we can write
\be\label{M5}
\kl N(t)\kr=\int d\psi(t)d\hat\psi (t)N(t)e^{\hat\psi (t)\psi(t)}
\hat g\big (\hat\psi (t)\big)g\big (\psi(t)\big),
\ee
with 
\ba\label{M6}
g\big (\psi(t)\big)=Z^{-1}_<\int \cD\psi(t'<t)\cD\hat\psi (t'<t)\hat T
\{e^{-S_<}\}g_{in},\nn\\
\hat g\big(\hat\psi (t)\big)=Z^{-1}_>\int \cD\psi(t'>t)\cD\hat\psi (t'>t)\hat g_f
\hat T\{e^{-S_>}\},
\ea
and $Z_<Z_>=Z$. This procedure integrates out the future and the past and we are left in eq. \eqref{M5} with a Grassmann integral involving only $\psi(t)$ and $\hat\psi (t)$ at a given time $t$. For a given $g_{in}$ we choose $\hat g_f$ such that $\hat g(\hat\psi (t))$ according to eq. \eqref{M6} is the conjugate element to $g\big(\psi(t)\big)$. Then eq. \eqref{M5} coincides with eq. \eqref{V11}. We will discuss this issue in more detail below and in sect. \ref{conjugategrassmann}. 

With this choice of $\hat g_f$ we have indeed expressed $\kl N(t)\kr$ in terms of the wave function $g(t)$
\be\label{73A}
\kl N(t)\kr =\int{D}\psi\hat g (t)N(t)g(t).
\ee
Both $g(t)$ and $\hat g (t)$ are explicitly constructed from the functional integral by eq. \eqref{M6}, defining the maps $G\to g(t)$ and $G \to \hat g (t)$. Our procedure easily generalizes to a more general form of the action and to more than one species $\psi_\alpha(t)$, provided the decomposition \eqref{M3}, \eqref{M4} is possible. If the action is normalized and contains only terms with an even number of Grassmann variables this expression simplifies
\ba\label{73B}
g(t)&=&\int_{t'<t}{\cal D}\psi{\cal D}\hat\psi  e^{-S_<}g_{in},\nn\\
\hat g (t)&=&\int_{t'>t}{\cal D}\psi{\cal D}\hat\psi \hat g_fe^{-S_>},
\ea
with an obvious meaning of the functional integrals.

\bigskip\noindent
{\bf 3. Time evolution}

\medskip\noindent
Let us next investigate the time evolution of $g(t)$. From the definition \eqref{M6} we can obtain $g(t+\epsilon)$ by adding an integral over the variables $\psi(t)$ and $\hat\psi(t)$, 
\ba\label{J1}
\partial_t g(t)&=&\frac{1}{\epsilon}\big(g(t+\epsilon)-g(t)\big)\nn\\
&=&\frac{1}{\epsilon}\Big\{\int d\psi(t)d\hat\psi (t)\exp \big\{-L(t)\big\}
g\big (\psi(t)\big)\nn\\
&&-g\big(\psi(t+\epsilon)\big)\Big\}.
\ea
Here we consider $g(t)=a(t)+b(t)\psi$ for a fixed Grassmann variable that we take as $\psi=\psi(t+\epsilon)$. We generalize $L(t)$ in eq. \eqref{M1} to
\be\label{J2}
L(t)=\hat\psi (t)\big (\psi(t+\epsilon)-\psi(t)\big)+i\epsilon {H}(t),
\ee
where $H(t)$ depends on $\psi(t+\epsilon)$ and $\hat \psi(t)$. For a general Hamiltonian $H[\hat{\psi}(t), \psi(t+\epsilon)]$ the Grassmann algebra is extended to a complex Grassmann algebra. For our particular example, however, $H$ is purely imaginary, such that $L$ remains an element of a real Grassmann algebra. Inserting the general expression \eqref{J2} into eq. \eqref{J1} one obtains
\ba\label{J3}
\partial_tg(t)&=&\frac{1}{\epsilon}\Big\{\int d\psi(t)d\hat\psi (t)
\exp(-i\epsilon{H})g_s(t)
-g\big(\psi(t+\epsilon)\big)\Big\}\nn\\
&=&-i\int d\psi(t)d\hat\psi (t){H}\big [\hat\psi (t),\psi(t+\epsilon)\big] g_s(t),\nn\\
g_s(t)&=&\exp \Big\{\hat\psi (t)\big (\psi(t)-\psi(t+\epsilon)\big)\Big\}g\big (\psi(t)\big).
\ea
This equation is exact for $H$ linear in $\psi,\hat\psi $ and gets corrections which vanish for $\epsilon\to 0$ in the more general case. 

In order to show that eq. \eqref{J3} describes a unitary evolution law we next eliminate the conjugate Grassmann variable $\hat\psi $, reversing the steps in the construction of sect. \ref{ConjugateGrassmannvariables}. Using the identity
\ba\label{J4}
\int d\psi(t)d\hat\psi (t)\hat\psi (t)g_s(t)
=\frac{\partial}{\partial\psi(t+\epsilon)}
\int d\psi(t)d\hat\psi (t) g_s(t)\nn\\
\ea
we get the evolution equation
\be\label{J5}
\partial_tg(t)=-i{\cal H}\left[\frac{\partial}{\partial\psi},\psi\right]g(t)
\ee
and identify ${\cal H}$ with the Grassmann-Hamilton operator. It obtains from $H$ by replacing $\hat\psi\to\partial/\partial\psi$, where the ordering puts $\partial/\partial\psi$ to the left of $\psi$. For our example one has
\ba\label{J6}
{H}\big[\hat\psi(t+\epsilon),\psi (t)\big]&=&i\omega
\big(\psi(t+\epsilon)-\hat\psi (t)\big),\nn\\
{\cal H}\left[\frac{\partial}{\partial\psi},\psi\right]&=&i\omega
\left(\psi-\frac{\partial}{\partial \psi}\right).
\ea
Eq. \eqref{J5} realizes the unitary time evolution \eqref{113B} and can therefore be associated to a unitary evolution law for classical probabilities. We will discuss in sect. \ref{cptsqm} a unitary time evolution of classical probabilities which realizes the Schr\"odinger equation
\be\label{94A}
\partial_t g(t)=\omega\left(\psi-\frac{\partial}{\partial\psi}\right)g(t).
\ee

Again, this construction is easily generalized. First one extracts from the action the Hamiltonian from eq. \eqref{J2}, with a generalization of the first term by a summation over species. Once $H$ is ordered such that all $\hat\psi _\alpha$-factors are to the left of the $\psi_\alpha$-factors one replaces $\hat\psi _\alpha\to \frac{\partial}{\partial\psi_\alpha}$ in order to obtain ${\cal H}$ from $H$. 

What remains to be shown is that the time evolution of $g(t)$ and $\hat g (t)$ as defined according to eq. \eqref{M6} or \eqref{73B}, is compatible with $\hat g (t)$ being the conjugate of $g(t)$ for all $t$. In other words, we want to establish that for $g(t)=\sum_\tau c_\tau(t)g_\tau\big(\psi(t)\big)$ one obtains $\hat g (t)=\sum_\tau c^*_\tau(t)\hat g _\tau\big(\hat\psi (t)\big)$ for all  $t$, provided this relation holds for some time $t_0$. We first note the identity
\be\label{80A}
{\cal N}_Z(t)=\int{D}\psi\hat g (t)g(t)=Z^{-1}
\int {\cal D}\psi(t){\cal D}\hat\psi (t)\hat g_f e^{-S}g_{in}=1.
\ee
If we define $\hat g (t)$ by eq. \eqref{M6} the time evolution preserves ${\cal N}_Z$ independently of $\hat g (t)$ being conjugate to $g(t)$ or not. If $g(t_{in})$ is normalized with 
${\cal N}_Z(t_{in})=1$ we are guaranteed that $Z=1$ without the need to normalize the action by the shift \eqref{M1}. This corresponds to the observation that we can obtain $Z=1$ by an appropriate rescaling of $g_{in}$ and $\hat g_f$ in eq. \eqref{M1A} or \eqref{80A}. 

Furthermore, if $\hat g (t)$ is conjugate to $g(t)$ for all $t$ the norm
\be\label{80AA}
{\cal N}_p(t)=\sum_\tau|c_\tau(t)|^2=\sum_\tau p_\tau(t)
\ee 
and ${\cal N}_Z(t)$ coincide. In this case we can guarantee $Z=1$ by imposing ${\cal N}_p(t_{in})=1$. The condition ${\cal N}_p(t_{in})$ does not involve any functional integral and can be formulated purely as a normalization condition for the coefficients $c_\tau(t_{in})$. 

We will show below that eq. \eqref{M6} implies 
\be\label{80B}
\partial_tg(t)=-i{\cal H}(t)~,~\partial_t\hat g (t)=i\hat{\cal H}^T\hat g (t),
\ee
where $\hat{\cal H}^T$ is related to ${\cal H}$ by the property that for arbitrary Grassmann elements $\hat g [\psi]$ and $f[\psi]$ one has the relation
\be\label{80C}
\int {D}\psi\hat{\cal H}^T\hat g f=\int {D}\psi\hat g  {\cal H} f.
\ee
For $\hat g (t)$ defined by eq. \eqref{M6} (and not necessarily conjugate to $g(t)$) we use the expansions
\be\label{80D}
g=\sum_\tau c_\tau (t)g_\tau~,~\hat g =\sum_\tau\hat c_\tau(t)\hat g _\tau.
\ee
Then eq. \eqref{80B} implies for the time evolution of the coefficients
\be\label{80E}
\partial_t c_\tau=\sum_\rho K_{\tau\rho}c_\rho~,~\partial_t\hat c_\tau=-\sum_\rho\hat c_\rho K_{\rho\tau}.
\ee
This follows from the relation
\ba\label{80F}
\partial_t\hat c_\tau&=&\int{D}\psi\partial_t\hat g  g_\tau=i\int {D}\psi
\hat{\cal H}^T\hat g  g_\tau\nn\\
&=&i\int D\psi\hat g {\cal H} g_\tau=-\int D\psi\hat g \sum_\rho g_\rho K_{\rho\tau}.
\ea
We note that eq. \eqref{80D} indeed preserves ${\cal N}_Z=\sum_\tau\hat c_\tau c_\tau$. In order to establish eq. \eqref{80B} we employ
\ba\label{80G}
\partial_t\hat g (t)&=&-\frac{1}{\epsilon}\big(\hat g (t-\epsilon)-
\hat g(t)\big)=\partial_t\hat g \big(\hat\psi (t-\epsilon)\big)\nn\\
&=&i \int d\psi(t)d\hat\psi (t)\hat g \big(\hat\psi (t)\big)\\
&&\exp \Big\{\big(\hat\psi (t)-\hat\psi (t-\epsilon)\big)\psi(t)\Big\}H(t-\epsilon).\nn
\ea
For an arbitrary element $f\big(\psi(t-\epsilon)\big)$ this yields indeed
\ba\label{80H}
&&\int D\psi(t-\epsilon)\partial_t\hat g \big(\hat\psi (t-\epsilon)f\big(\psi(t-\epsilon)\big)\nn\\
&&=i\int{D}\psi(t-\epsilon)\hat g \big(\hat\psi (t-\epsilon)\big){\cal H} f\big(\psi(t-\epsilon)\big).
\ea

The condition that $\hat g (t)$ remains conjugate to $g(t)$ for all $t$ amounts to $c^*_\tau(t)=\hat c_\tau(t)$. From $\partial_t c^*_\tau=\sum_\rho K^*_{\tau\rho}c^*_\rho$ we infer the condition
\be\label{80I}
K^*_{\tau\rho}=-K_{\rho\tau}.
\ee
In terms of the Hamiltonian $H=iK$ this requires hermiticity of $H$, 
\be\label{80J}
H^\dagger=H.
\ee
Only for an hermitean Hamiltonian the Grassmann functional integral can describe the time evolution of classical probabilities with constant normalization $\sum_\tau p_\tau (t)=1$. For a real Grassmann algebra $K$ is a real matrix and must be antisymmetric. Eq. \eqref{80I} is obeyed for our example. We observe that the evolutions \eqref{80B} or \eqref{80E} are invertible if $\exp \big\{K(t-t_0)\big\}$ is an invertible matrix, which holds, in particular, for $K=-iH,H^\dagger=H$. 

\bigskip\noindent
{\bf 4. Solution of evolution equation}

The formal solution of the evolution equation \eqref{J5} is given by the central identity \eqref{M6} or \eqref{73B}, provided $\hat g_f=\hat g (t_f)$. Here the initial condition for the differential equation \eqref{80D} can be specified at $t_{in}$ by $g_{in}=g(t_{in})$. Alternatively, we can specify a particular solution by indicating $g(t_0)$ for some arbitrary time $t_0$. Then $g_{in}=g(t_{in})$ and $g(t_f)$, and therefore also $\hat g_f=\hat g (t_f)$, can be computed from the evolution equation. (For $\hat g_f$ and $g_{in}$ not related to the solution of the evolution equation the functional integral \eqref{M1A} yields a transition matrix instead of $Z$.)

The identity \eqref{M6} or \eqref{73B} can be interpreted in different useful ways. We may introduce the Grassmann element
\be\label{117Aa}
G_<=Z^{-1}_<\hat T\{e^{-S_<}\}g_{in},
\ee
which depends on the Grassmann variables $\psi(t'<t),$ $\hat\psi (t'<t)$ and $\psi(t)$. We can then interprete $g\big(\psi(t)\big)$ as an integration over ``unobservable variables''
\be\label{117B}
g\big(\psi(t)\big)=\int \cD\psi(t'<t)\cD\hat\psi (t'<t) G_<.
\ee
This underlines the character of the local (in time) wave function and probability as a subsystem where degrees of freedom of a ``total system'' are integrated out \cite{3}. The total system may be associated with a probability distribution for all times, where ``probabilistic time'' appears only as a particular ordering structure \cite{3A}. ``Unobservable'' means in this context that observations are performed only with local observables. A similar identity holds for $\hat g \big(\hat\psi (t)\big)$, 
\ba\label{117}
&&G_>=Z_>^{-1}\hat g_f\hat T\{e^{-S_>}\}\nn\\
&&\hat g \big(\hat\psi (t)\big)=\int \cD\psi(t'>t)\cD\hat\psi (t'>t)G_>,
\ea
where $G_>$ depends on $\psi(t'>t),\hat\psi (t'>t)$ and $\hat\psi (t)$.

Using eq. \eqref{M4}  we can decompose $Z_<\hat T\{e^{-S_<}\}$ into a product of factors
\ba\label{117D}
Z^{-1}_<\hat T\{e^{-S_<}\}&=&\hat T\left\{\prod_n e^{-s_n}\right\}\nn\\
e^{-s_n}&=&z^{-1}e^{-L(t_n)},
\ea
where the product is over all $n$ with $t_{in}\leq t_n<t$ and $\prod_nz^{-1}=Z^{-1}_<$. Each individual factor $s_n$ depends on $\hat\psi (t_n)$, $\psi(t_n)$ and $\psi(t_{n+1})=\psi(t_n+\epsilon)$. We define formally the Grassmann element
\be\label{117E}
u(t_{n+1},t_n)=\int d\hat\psi (t_n)e^{-s_n}
\ee
which depends on $\psi(t_n)$ and $\psi(t_{n+1})$. The operator $\bar U$ introduces a new multiplication law which includes the Grassmann integral $\int d\psi(t_n)$
\ba\label{117F}
&&\bar U(t_{n+1},t_n)\circ g\big(\psi(t_n)\big)\equiv
\int d\psi(t_n)u(t_{n+1},t_n)g\left(\psi(t_n)\right)\nn\\
&&=\int d\psi(t_n) d\hat\psi (t_n)
\Big\{e^{-s_n}g\big(\psi(t_n)\big)\Big\},
\ea
such that $\bar U(t_{n+1},t_n)\circ g(\big(\psi(t_n)\big)$ depends on $\psi(t_{n+1})$. Thus $\bar U$ can be viewed also as an operator that maps local wave functions at $t_n$ onto local wave functions at $t_{n+1}$. This allows us to write $g\big(\psi(t)\big)$ in the form
\be\label{117G}
g(t)=g\big(\psi(t)\big)=\prod_n\big(\bar U(t_{n+1},t_n)\circ\big) g\big(\psi(t_{in})\big),
\ee
where the operator product is defined in analogy to eq. \eqref{117F}, ordered such that larger time arguments are on the left.

\section{Classical probabilities for two-state quantum mechanics}
\label{cptsqm}
In this section we specify of an explicit unitary evolution law for classical probabilities that realizes the evolution equation \eqref{94A}. It reads
\ba\label{61}
p_1(t)=\cos^2(\omega t)p_1(0)+\sin^2(\omega t)p_0(0)\nn\\
-2s\cos (\omega t)\sin(\omega t)
\sqrt{p_1(0)p_0(0)},
\ea
where we set $t_0=0$ and $s=\pm 1$. We note $0\le p_1(t)\le 1$ and define $p_0(t)=1-p_1(t)$, such that the probabilities for the two states $0$, $1$ are normalized, $p_0(t)+p_1(t)=1$. If we interprete $p_1(t)$ as the probability for the occupied state, and $p_0(t)$ as the one for the empty state, the mean occupation number obeys
\be\label{61A}
\kl N(t)\kr=p_1(t).
\ee

The rotation matrix $R$ in eq. \eqref{A} can be written as a real unitary matrix $U$
\be\label{63}
R=\left(\begin{array}{ll}
\cos \omega t,&-\sin \omega t\\\sin\omega t,&\cos \omega t
\end{array}\right)=U=\exp (-i\omega\tau_2 t).
\ee
This corresponds to two-state quantum mechanics with a hermitean Hamiltonian $H=\omega\tau_2$ and a real two-component initial wave function $\varphi(0)=\big(q_1(0),q_0(0)\big)=(\sqrt{p_1(0)},s\sqrt{p_0(0)})$. 
Using the standard quantum formalism for a two-component wave function one has
\ba\label{117A}
\varphi(t)=\left(\begin{array}{l}{q_1(t)}\\{q_0(t)}\end{array}\right)~,~\partial_t\varphi(t)=-iH\varphi(t)~,
~H=\omega\tau_2.
\ea
The wave function $\varphi(t)=U(t,0)\varphi(0)$ remains real in the course of the evolution. Evaluating the expectation value of the operator $\hat N=\frac12(1+\tau_3)$, which is associated to the occupation number observable,
\ba\label{64}
\kl N(t)\kr=\kl\varphi(t)|\hat N|\varphi(t)\kr=
\kl\varphi(0)U^\dagger(t,0)|\hat N|U(t,0)\varphi(0)\kr,\nn\\
\ea
one recovers eq. \eqref{61A}.

The Grassmann algebra (at a given $t$) contains only two basis elements, $\{g_\tau\}=\{g_1,g_0\}=\{1,\psi\}$ and the state is represented by
\ba\label{65}
g(t)&=&q_1(t)+q_0(t)\psi,\nn\\
q_1(t)&=&\cos(\omega t)\sqrt{p_1(0)}-\sin(\omega t)s\sqrt{p_0(0)},\nn\\
q_0(t)&=&\sin (\omega t)\sqrt{p_1(0)}+\cos(\omega t)s\sqrt{p_0(0)},\nn\\
p_1(t)&=&q_1(t)^2~,~p_0(t)=q_0(t)^2.
\ea
The Grassmann evolution operator
\ba\label{66}
{\cal U}(t)&=&\cos(\omega t)+\sin (\omega t)\left(\psi-\frac{\partial}{\partial \psi}\right)\nn\\
&=&\cos(\omega t)+\sin (\omega t)(a-a^\dagger)
\ea
describes the time evolution
\be\label{67}
g(t)={\cal U}(t)g(0).
\ee
The conjugate basis elements read $\{\tilde g_\tau\}=\{\psi,1\}$, with
\ba\label{68}
\tilde g(t)&=&q_0(t)+q_1(t)\psi.
\ea

Using $\left(\psi-\frac{\partial}{\partial\psi}\right)^2=-1$ we can write
\be\label{70}
{\cal U}(t)=\exp\left\{\omega
\left(\psi-\frac{\partial}{\partial\psi}\right)t\right\}.
\ee
This can be cast into a Hamiltonian form similar to quantum mechanics,
\ba\label{71}
i\partial_t{\cal U}&=&{\cal H}{\cal U}~,~{\cal U}=\exp (-i{\cal H}t),\nn\\
{\cal H}&=&i\omega
\left(\psi-\frac{\partial}{\partial\psi}\right)=i\omega(a-a^\dagger),
\ea
with
\ba\label{72}
a=\left(\begin{array}{l}0,0\\1,0\end{array}\right)~,~
a^\dagger=\left(\begin{array}{l}0,1\\0,0\end{array}\right).
\ea
We recover the evolution \eqref{J5}, \eqref{J6}, \eqref{94A}, demonstrating that the functional integral defined by eq. \eqref{M1} indeed describes the time evolution of classical probabilities.

In the Grassmann formulation the occupation number is represented by the Grassmann operator
\be\label{73}
{\cal N}=\frac{\partial}{\partial\psi}\psi=a^\dagger a.
\ee
It translates to the quantum operator $\hat N=(1+\tau_3)/2$, with
\ba\label{74}
\kl N(t)\kr &=&\int d\psi\tilde g(t)\frac{\partial}{\partial\psi}\psi g(t)\\
&=&\kl\varphi(t)|\hat N|\varphi(t)\kr
=\varphi^\dagger(t)\frac{1+\tau_3}{2}\varphi(t).\nn
\ea
At this stage we do not yet have a full description of two-state quantum mechanics for the non-commuting spin observables in the different directions, even though a non-commuting operator appears in the form of ${\cal H}$ in eq. \eqref{71}, cf. ref. \cite{3A}. The implementation of non-commuting observables within a classical statistical ensemble, including spin, is discussed in detail in ref. \cite{CW??}.

\section{Grassmann functional integral from classical probabilities}
\label{conjugategrassmann}

The inverse road from classical probabilities $\{p_\tau \}$ to a Grassmann functional integral is also open. In this section we start with a classical statistical ensemble for two states whose probability distribution follows the time evolution \eqref{61}. Using the associated wave function $q_\tau(t)$ and the associated Grassmann element $g(t)$ given by eq. \eqref{65}, we realize that this ensemble can be interpreted as two-state quantum mechanics. We employ the conjugate Grassmann variables in order to derive the associated functional integral. This reverses the steps of sect. \ref{quantumwavefunction} and demonstrates explicitly how a Grassmann functional integral obtains from a given time evolution of classical probabilities. The construction of the functional integral follows the standard concept of decomposing the evolution operator into a product of infinitesimal factors \cite{Zin}. We display it here in our formulation in order to show how the functional integral arises naturally if we want to express ${\cal U}(t)$ in eq. \eqref{70} in terms of conjugate Grassmann variables. In particular, we will not need the often employed concept of coherent states. The formalism of this section is easily generalized to arbitrary real or complex wave functions for states characterized by occupation numbers $[n_\beta]$ with unitary time evolution according to eq. \eqref{67}.

In terms of the conjugate Grassmann variables $\hat\psi $ the basis elements conjugate to $g_\tau=\{1,\psi\}$ are given by $\{\hat g _\tau\}=\{1,\hat\psi \}$, and the Grassmann element conjugate to $g(t)=q_1(t)+q_0(t)\psi$ reads 
\be\label{133A}
\hat g (t)=q_1(t)+q_0(t)\hat\psi .
\ee
We want to implement the time evolution operator \eqref{70} in the formulation with conjugate Grassmann variables $\hat\psi (t)$. In order to perform the necessary normal ordering of the factors $\psi$ and $\partial/\partial\psi$ in the exponential \eqref{70} we proceed by the standard methods of a formulation of a path integral. This will bring us back to the functional integral \eqref{M2Aa} and generalizations thereof. The central idea is a decomposition of ${\cal U}(t,t_0)$ into a product of operators with infinitesimal time steps ${\cal U}(t_{n+1},t_n)$, $t_{n+1}-t_n=\epsilon\to 0$, such that the appropriate ordering for 
${\cal U}(t_{n+1},t_n)$ becomes trivial up to negligible corrections in higher orders in $\epsilon$. 

\bigskip\noindent
{\bf 1. Factorization of evolution operator}

\medskip\noindent
We first proceed to a split of the evolution operator into factors for subintervals in time. These intervals can be taken infinitesimally small. For an arbitrary operator ${\cal F}[\psi]=a+b\psi+c\frac{\partial}{\partial\psi}+d\frac{\partial}{\partial\psi}\psi$ and an arbitrary element $g(\psi)=g+h\psi$ one has the identity
\ba\label{T1}
&&\int d\psi~e^{\hat\psi \psi}{\cal F}[\psi]g(\psi)\\
&&=\int d\psi'~e^{\hat\psi \psi'}{\cal F}[\psi']\int d\psi d\hat\psi  '~
e^{\hat\psi '(\psi-\psi')}g(\psi),\nn
\ea
with ${\cal F}[\psi']=a+b\psi'+c\frac{\partial}{\partial \psi'}+d\frac{\partial}{\partial\psi'}\psi'$. This can be used for a decomposition of the product of two operators
\ba\label{T2}
&&\int d\psi e^{\hat\psi \psi}{\cal F}_1[\psi]{\cal F}_2[\psi]g(\psi)\\
&&=\int d\psi' e^{\hat\psi \psi'}{\cal F}_1[\psi']\int d\psi d\hat\psi ' 
e^{\hat\psi '(\psi-\psi')}{\cal F}_2[\psi]g(\psi).\nn
\ea
(The  product is here the standard product of Grassmann operators and should not be confounded with the product $\circ$ used in eqs. \eqref{117F}, \eqref{117G}.)
In particular, we can employ (for arbitrary ${\cal B}[\psi]$)
\ba\label{T3}
{\cal B}{\cal U}(t,t_0)&=&{\cal F}_1{\cal F}_2~,~{\cal F}_1={\cal B}{\cal U}(t,t_1),\nn\\
{\cal F}_2&=&{\cal U}(t_1,t_0),
\ea
and identify $\hat\psi '=\hat\psi (t_0),\psi'=\psi(t_1),\psi=\psi(t_0)$, such that 
\ba\label{T4}
&&\int d\psi~e^{\hat\psi \psi}\big({\cal B}{\cal U}(t,t_0)\big)[\psi]g(\psi)
=\int d\psi(t_1)e^{\hat\psi \psi(t_1)}\nn\\
&&\quad\big ({\cal B}{\cal U}(t,t_1)\big)[\psi(t_1)]
\int d \psi(t_0)d\hat\psi (t_0)
e^{-\hat\psi (t_0)
\big (\psi(t_1)-\psi(t_0)\big)}\nn\\
&&\quad {\cal U}(t_1,t_0)[\psi(t_0)]g\big (\psi(t_0)\big).
\ea
For $t_1-t_0=\epsilon$ the integral $\int d\psi(t_0)d\hat\psi (t_0)$ involves in the integrand only an infinitesimal evolution operator that can be evaluated in lowest non-trivial order in $\epsilon$. 

This procedure can be repeated for ${\cal B}{\cal U}(t,t_1)$. We end with with a sequence of simple Grassmann integrals over infinitesimal evolution operators
\ba\label{T5}
&&\int d\psi~e^{\hat\psi \psi}\big({\cal B}{\cal U}(t,t_0)\big)[\psi]g(\psi)
=\int d\psi(t)~e^{\hat\psi \psi(t)}
{\cal B}[\psi(t)]\nn\\
&&\quad\prod^{N-1}_{n=0}\Big\{\int d\psi(t_n)d\hat\psi (t_n)~
e^{-\hat\psi (t_n)\big(\psi(t_{n+1})-\psi(t_n)\big)}\nn\\
&&\quad{\cal U}(t_{n+1},t_n)[\psi(t_n)]\Big\}g\big (\psi(t_0)\big).
\ea
Here $t_N=t$ and factors with larger $t_n$ are to the left of factors with smaller $t_n$. We recognize in the first factor the ``shift operator''
\ba\label{144A}
\bar S(t_{n+1},t_n)&=&\int d\hat\psi  (t_n)
e^{-\hat\psi (t_n)\big(\psi(t_{n+1})-\psi(t_n)\big)}\nn\\
&=&\delta\big(\psi(t_n)-\psi(t_{n+1})\big)
\ea
with multiplication law
\ba\label{144B}
&&\bar S(t_{n+1},t_n)\circ
g\big(\psi(t_n)\big)\equiv \\
&&\int d\psi(t_n)\bar S(t_{n+1},t_n)
g\big(\psi(t_n)\big)
=g\big(\psi(t_{n+1})\big).\nn
\ea
Its role is to shift the argument of $g$, and we recognize in eq. \eqref{T5} the product $\circ$ with factors $\bar S(t_{n+1},t_n){\cal U}\tn$, which first apply the Grassmann evolution operator ${\cal U}$ and subsequently shift the argument. For ${\cal B}=1$ this yields the factorization of ${\cal U}$,
\ba\label{109A}
&&\int d\psi(t_0)e^{\hat\psi \psi(t_0)}
{\cal U}(t,t_0)\big[\psi(t_0)\big]g\big(\psi(t_0)\big)=
\int d\psi(t)e^{\hat\psi \psi(t)}\nn\\
&&\quad\times  \prod^{N-1}_{n=0}\Big(\bar S(t_{n+1},t_n)\circ 
{\cal U}(t_{n+1},t_n)\big[\psi(t_n)\big]\Big)g\big(\psi(t_0)\big).\nn\\
\ea

\bigskip\noindent
{\bf 2. Hamilton operator}

In the limit $\epsilon=t_{n+1}-t_n\to 0$ it is sufficient to consider the lowest order in $\epsilon$,
\be\label{T6}
{\cal U}(t_{n+1},t_n)[\psi(t_n)]=1-i{\cal H}[\partial/\partial\psi(t_n),\psi(t_n)]\epsilon.
\ee
Since ${\cal H}$ is multiplied by a $\delta$-function from $\bar S\tn$ we can replace the argument $\psi(t_n)$ by $\psi(t_{n+1})$. Furthermore, if ${\cal H}$ is normal ordered, as for eq. \eqref{71}, we can replace $\partial/\partial\psi(t_n)$ by $\hat\psi (t_n)$. In total, we perform the replacement
\ba\label{T7}
&&{\cal U}(t_{n+1},t_n)[\psi(t_n)]\to 1-i{H}[\hat\psi (t_n),\psi(t_{n+1})]\epsilon\nn\\
&&\hspace{1.4cm}=\exp \Big\{-i{H}[\hat\psi (t_n),\psi(t_{n+1})]\epsilon\Big\}.
\ea
For the particular example of two quantum mechanics this yields
\be\label{T7a}
{\cal U}\tn\to 
\exp[\epsilon\omega\big (\psi(t_{n+1})-\hat\psi (t_n)].
\ee

We can now use eq. \eqref{109A} for a computation of $g(t)$, 
\ba\label{146A}
&\int& d\psi(t_0) e^{\hat\psi  \psi(t_0)}{\cal U}(t,t_0)[\psi(t_0)]g(t_0)\nn\\
&&=\int d\psi(t)e^{\hat\psi  \psi(t)}g(t).
\ea
With
\ba\label{114Y}
\bar U\tn  
g\big(\psi(t_n)\big)=\int d\psi(t_n){u}(t_{n+1},t_n)g\big(\psi(t_n)\big)\nn\\
\ea
and
\ba\label{114Z}
u(t_{n+1},t_n)&=&\int d\hat\psi (t_ n)
\exp \Big\{-i\epsilon H\big(\hat\psi (t_n),\psi(t_{n+1}\big)\Big\}\nn\\
&&\exp \Big\{\hat\psi (t_n)\big(\psi(t_n)-\psi\big(t_{n+1})\big)\Big\}
\ea
one obtains
\ba\label{114X}
g(t)&=&\int d\psi(t_0)d\hat\psi (t_0)
e^{\hat\psi (t_0)\big(\psi(t_0)-\psi(t)\big)}\nn\\
&&{\cal U}(t,t_0)\big[\psi(t_0)\big]g(t_0)\nn\\
&=&\prod^{t-\epsilon}_{t_n=t_0}\big(\bar U(t_{n+1},t_n)\circ \big)g(t_0).
\ea
Here $g(t)$ depends on $\psi(t)$ and $g(t_0)$ depends on $\psi(t_0)$, while multiplication with $\bar U(t_{n+1},t_n)$ involves a shift from $\psi(t_n)$ to $\psi(t_{n+1})$ where the product $\circ$ is employed. We recognize eq. \eqref{117G}. The derivation of eq. \eqref{146A} has not used any specific properties of ${\cal U}$ except the existence of an expansion \eqref{T6} in lowest order in $\epsilon$. The results of this section can be used for arbitrary ${\cal U}$ or ${\cal H}$, with trivial generalization to $g$ constructed from an arbitrary number of Grassmann variables.

\bigskip\noindent
{\bf 3. Functional integral for observables}

\medskip\noindent
We want to express the expectation value of some observable $\kl A(t)\kr$ in terms of the initial wave function $g\big(\psi(t_0)\big)$. According to eqs. \eqref{29}, \eqref{42a} we write
\ba\label{114A}
\kl A(t)\kr&=&\int d\psi(t)\tilde g(t){\cal A} g(t)\nn\\
&=&\int d \psi(t_0)\tilde g(t_0)\tilde {\cal U}(t,t_0){\cal A}{\cal U}(t,t_0)g(t_0),
\ea
and express with eq. \eqref{V2}
\ba\label{114B}
\tilde g(t_0)=\int d\hat\psi \hat g (\hat\psi ) e^{\hat\psi \psi(t_0)}.
\ea
We therefore need an expression for $\tilde{\cal U}(t,t_0){\cal A}(t){\cal U}(t,t_0)$. For this purpose we consider in eq. \eqref{T3}
\be\label{T8}
{\cal B}[\psi(t)]=\big (\tilde {\cal U}(t_0,t){\cal A}\big)[\psi(t)].
\ee
We can continue the factorization procedure and end with the main result of this section
\ba\label{T9}
&&\kl A(t)\kr=\int d\psi d\hat\psi \hat g (\hat\psi )~e^{\hat\psi \psi}
\big (\tcu(t,t_0){\cal A}{\cal U}(t,t_0)\big)[\psi]g(\psi)\nn\\
&&=\prod^{2N}_{k=0}\Big[\int d\psi(t_k)d\hat\psi (t_k)\Big]
\hat g \big (\hat\psi (t_{2N})\big )~e^{\hat\psi (t_{2N})\psi(t_{2N})}\nn\\
&&\quad \times\prod^{2N-1}_{m=N}\Big[\exp\big\{i{H}[\hat\psi (t_{m}),\psi(t_{m+1})]
(t_m-t_{m+1})\big\}\nn\\
&&\quad \times\exp \big\{-\hat\psi (t_m)\big (\psi(t_{m+1})-\psi(t_m)\big)\big\}\Big]
{A}[\hat\psi (t),\psi(t)]\nn\\
&&\quad\times\prod^{N-1}_{n=0}
\Big[\exp\big\{-i{H}[\hat\psi (t_n),\psi(t_{n+1})](t_{n+1}-t_n)\big\}\nn\\
&&\quad\times\exp \big\{-\hat\psi (t_n)\big(\psi(t_{n+1})-\psi(t_n)\big)\big\}\Big]
g\big (\psi(t_0)\big),
\ea
where $t=t_N$ and $\hat\psi =\hat\psi (t_{2N})$. The order of the factors is with increasing $n$ and $m$ from right to left. Counting the $t_m$ backwards, $t_{m+1}-t_m=-\epsilon$ we associate $t_{2N}=t_0$. For a given $t_n=t_0+n\epsilon$ we end then with two distinct Grassmann variables $\psi(t_n)$ and $\varphi(t_n)=\psi(t_{2N-n})$ (except for $\psi(t_N)$.) This amounts to the Schwinger-Keldish formalism in quantum field theory \cite{SK}. 

The structure of eq. \eqref{T9} can also be understood from the equivalent expression
\be\label{116A}
\kl A(t)\kr=\int d\psi(t)d\hat\psi (t)
e^{\hat\psi (t)\psi(t)}\hat g (t){A}\big[\hat\psi (t),\psi(t)\big]g(t),
\ee
where $g(t)$ is given by eq. \eqref{114X} and $\hat g (t)$ obeys
\be\label{116B}
\hat g (t)=\hat g (t_0)\prod^{t-\epsilon}_{t_n=t_0}\big(\circ \bar V(t_n,t_{n+1})\big).
\ee
In this expression $\hat g (t)$ depends on $\hat\varphi(t)=\hat\psi (t)$ and $\hat g (t_0)$ on $\hat\varphi(t_0)$, while multiplication from the right with $\bar V(t_n,t_{n+1})$ involves a shift from arguments $\hat\varphi(t_n)$ to $\hat\varphi(t_{n+1})$. The order of factors in eq. \eqref{116B} is now with smaller $t_n$ to the left. The product involves an integration over $\hat\varphi(t_n)$,
\be\label{116C}
\hat f\big(\hat\varphi(t_n)\big)\circ \bar V(t_n,t_{n+1})=
\int d\hat\varphi(t_n)\bar f\big(\hat\varphi(t_n)\big)v(t_n,t_{n+1}).
\ee
For a hermitean Hamiltonian one has
\ba\label{116D}
&&v(t_n,t_{n+1})=v\big(\bv(t_n),\bv\tnn\big)\nn\\
&&=\int d\varphi(t_n)\exp \Big\{\varphi(t_n)\big(\bv(t_n)-\bv(t_{n+1})\big)\Big\}\nn\\
&&\hspace{1.85cm}\exp \Big\{i\epsilon H\big(\bv\tnn,\varphi(t_n)\big)\Big\}.
\ea
Here $\bar f$ obtains from $\hat f$ by changing the sign of all odd products of Grassman variables. 

The fact that $\tcu(t,t_0)$ is an inverse of ${\cal U}(t,t_0)$ is reflected by the relation (for $\hat\varphi(t)\equiv \hat\psi(t)\big)$
\ba\label{116E}
&&\int d\psi(t)d\hat\psi (t)e^{\hat\psi (t)\psi(t)}\int d\hat\varphi(t-\epsilon)\bar f\big(\hat\varphi(t-\epsilon)\big)
v(t-\epsilon,t)\nn\\
&&\quad \times\int d\psi(t-\epsilon)u(t,t-\epsilon)g\big(\psi(t-\epsilon)\big)\nn\\
&& =\int d\psi(t-\epsilon)\int d\hat\varphi(t-\epsilon)
e^{\bv(t-\epsilon)\psi(t-\epsilon)}\nn\\
&&\quad \times \hat f\big(\hat\varphi(t-\epsilon)\big) g\big(\psi(t-\epsilon)\big).
\ea
For arbitrary $t'<t$ and arbitrary $g(t')=g\big(\psi(t')\big)$ and $\hat f(t')=\hat f\big(\bv(t')\big)$ one has
\ba\label{116F}
&&\int d\psi(t)d\hat\varphi(t)e^{\hat\varphi (t)\psi(t)}\hat f(t')\nn\\
&&~~ \circ \prod^{t-\epsilon}_{t_{n'}=t'}\bar V(t_{n'},t_{n'+1})
\prod^{t-\epsilon}_{t_n=t'}\bar U\tn \circ  g(t')\nn\\
&&=\int d\psi(t')d\bv(t')e^{\bv(t')\psi(t')}\hat f(t')g(t').
\ea

A convenient way to understand these relations relies on the familiar concept of transition amplitudes which can be defined in the Grassmann formalism as
\ba\label{116G}
\kl \hat f|g\kr=\int d\psi d\hat \psi e^{\hat\psi\psi}\hat f(\hat\psi )g(\psi)=
\int {D}\psi\hat f(\hat\psi )g(\psi),
\ea
with $\kl \hat g |g\kr=1$. In this language the relation \eqref{116E} reads
\ba\label{151A}
\kl \hat f(t)|g(t)\kr&=&\kl \hat f(t-\epsilon)\circ \bar V(\tep,t)|\bar U(t,\tep)
\circ g(\tep)\kr\nn\\
&=&\kl \hat f(\tep)|g(\tep)\kr,
\ea
while eq. \eqref{116F} can be expressed as 
\be\label{151B}
\kl \hat f(t)|g(t)\kr=\kl \hat f(t')|g(t')\kr.
\ee
For a Grassmann operator ${\cal F}_N(\partial/\partial\psi,\psi)$ we generalize eq. \eqref{V7}
\ba\label{116H}
&&\kl \hat f|{\cal F}_Ng\kr=\kl\hat f|{\cal F}_N|g\kr\nn\\
&&=\int {D}\psi \hat f(\hat\psi ){F}_N(\hat\psi ,\psi)g(\psi),
\ea
where it is understood that $\partial/\partial\psi$ - and $\hat\psi $-factors in ${\cal F}_N(\partial/\partial\psi,\psi)$ and ${F}_N(\hat\psi ,\psi)$ are to the left of $\psi$-factors.

\bigskip\noindent
{\bf 4. Functional integral with future time}

\medskip\noindent
The form of eq. \eqref{T9} is suggesting a different option, namely to count $t_m$ to the future, $t_{m+1}-t_m=\epsilon$, associating $t_{2N}=2t-t_0$. We will see below that the corresponding change of the sign of $\epsilon$ in the terms involving $H\big[\hat\psi (t_m),\psi(t_{m+1})\big]$ can be absorbed by a simple change of $\hat g \big(\hat\psi (t_{2N})\big)$. Instead of being conjugate to $g(t_0)$ it is  now conjugate to $g(t_{2N})=g(2t-t_0)$. In consequence, eq. \eqref{T9} yields the functional integral expression
\ba\label{T10}
&&\kl A(t)\kr=\int\cD\psi\cD\hat\psi (t)\hat g (t+s)
{A}[\hat\psi (t),\psi(t)]\nn\\
&&\hspace{1.5cm}\times \hat T\Big\{\exp\big\{-\sum_{t'}L(t')\big\}\Big\}
g(t-s),
\ea
with $\hat g (t+s)$ depending on $\hat\psi (t+s)$ and $g(t-s)$ involving $\psi(t-s)$, whereby
\be\label{218A}
L(t')=\hat\psi (t')\big(\psi(t'+\epsilon)-\psi(t')\big)+
i\epsilon H\big[\hat\psi (t'),\psi(t'+\epsilon)\big].
\ee
In eq. \eqref{T10} we integrate over all $\psi(t'),\hat\psi (t')$ and $t'$ covers $2N+1$ values in the interval $t-s\leq t'\leq t+s~,~s=N\epsilon$, with $L(t+s)=-\hat\psi (t+s)\psi(t+s)$. We have assumed that ${A}$ is a bosonic observable, containing only terms with an even number of $\psi,\hat\psi $. The time ordering $\hat T$ is only needed if ${H}$ contains terms with odd powers of $\psi,\hat\psi $, such that ${H}(t_a)$ and ${H}(t_b)$ do not commute. The action of $\hat T$ is to put factors $\exp -i\epsilon{H}(t')$ with larger $t'$-arguments to the left of factors with smaller $t'$ arguments. 

We recall that $\hat g (t+s)$ is conjugate to $g (t+s)$ and $g(t+s)$ is related to $g(t-s)$ by the evolution law. Therefore only one independent element of the Grassmann algebra specifies the state of the system. The expression \eqref{T10} equals the functional integral \eqref{M2Aa} (for $Z=1$) if we identify $g_{in}=g(t-s)$ and $\hat g_f=\hat g (t+s)$. This direct construction of the relation between $\hat g_f$ and $g_{in}$ coincides with the requirements for a consistent evolution of $g(t)$ and $\hat g (t)$ defined by eq. \eqref{M6} and discussed in sect. \ref{quantumwavefunction}. For our specific example we have derived the Grassmann representation for the simple time evolution for classical probabilities \eqref{61}. The formulation in terms of $\hat\psi $ and $\psi$ leads directly to the functional integral expression \eqref{M2Aa} in sect. \ref{quantumwavefunction}. 

What remains to be shown is that the functional integral \eqref{T9} with $t_{m+1}>t_m$ and $\hat g \big(\hat\psi (t_{2N}\big)=\hat g (t+s)$ indeed yields the expression \eqref{114A} or \eqref{116A}. For this procedure we write eq. \eqref{116A} in a form that can directly be  used in eq. \eqref{116F} $(t'=t_N,t=t_{2N})$, 
\ba\label{118A}
&&\kl A(t)\kr=\int d\psi(t_N)d\bv(t_N)e^{\bv(t_N)\psi(t_N)}
\hat g (t_N)({\cal A}g)(t_N)\nn\\
&&=\int d\psi(t_{2N})d\hat\psi (t_{2N})e^{\hat\psi (t_{2N})\psi(t_{2N})}
\hat g (t_N)\nn\\
&&\times\prod^{t_{2N-1}}_{t_{n'}=t_N}\bar V(t_{n'},t_{n'+1})
\prod^{t_{2N-1}}_{t_n=t_N}\bar U(t_{n+1},t_n)({\cal A}g)(t_N)
\ea
With
\be\label{118B}
\hat g (t_N)\prod^{t_{2N-1}}_{t_{n'}=t_N}\bar V(t_{n'},t_{n'+1})=\hat g (t_{2N}),
\ee
and replacing again ${\cal A}\big[\partial/\partial\psi(t_N),\psi(t_N)\big]\to 
{A}\big[\hat\psi (t_N),\psi(t_N)\big]$, we find indeed eq. \eqref{T9} with $t_{m+1}>t_m$ and $\hat g \big (\hat\psi (t_{2N}\big)=\hat g (t_{2N})=\hat g (t+s)$.

\section{Two component spinor and complex structure}
\label{Twocomponentspinor}

Our first example has involved a real wave function and a real Grassmann algebra. In this section we consider a simple system that admits a complex structure. As a consequence, we can equivalently use a complex wave function and a complex Grassmann algebra for a description of this system. The opposite mapping from the complex description to the real one is rather trivial, since complex numbers can always be expressed in terms of a pair of real numbers. 

\bigskip\noindent
{\bf 1. Four state quantum mechanics}

\medskip\noindent
Our example describes a four state system, $\tau=1\dots{\cal N}_s$, ${\cal N}_s=4$, $B=2$. The time evolution of the classical probabilities is given by
\ba\label{P5A}
p_1(t)&=&p_{1,0},\nn\\
p_2(t)&=&\cos^2(\omega t)p_{20}+\sin^2(\omega t)p_{30}\nn\\
&&-2\cos(\omega t)\sin(\omega t)\sqrt{p_{20}p_{30}},\nn\\
p_3(t)&=&\cos^2(\omega t)p_{30}+\sin^2(\omega t)p_{20}\nn\\
&&+2\cos(\omega t)
\sin (\omega t)\sqrt{p_{20}p_{30}},\nn\\
p_4(t)&=&p_{40},
\ea
where $p_{\tau,0}$ are the initial probabilities at $t=0$, normalized by $p_{10}+p_{20}+p_{30}+p_{40}=1$. This evolution can be described by a differential equation for the real wave function with components 
$\{q_\tau\},p_\tau=q^2_\tau$,
\be\label{P3A}
\partial_tq_1=\partial_tq_4=0~,~\partial_tq_2=-\omega q_3~,~\partial_tq_3=\omega q_2.
\ee
We observe the close analogy with sect. \ref{quantumwavefunction}, with $q_2$ and $q_3$ playing the role of $q_1$ and $q_0$ in eq. \eqref{117A}.

However, we have now a four-state system with other possible observables. It admits a Grassmann representation different from sect. \ref{quantumwavefunction}. A suitable basis for a real Grassmann algebra consists of $\{g_\tau\}=(1,\psi_1,\psi_2,\psi_1\psi_2)$. In this basis the Hamiltonian for our example can be written as
\be\label{P7A}
{\cal H}=-i\omega\frac{\partial}{\partial \psi_1}\psi_2+i\omega\frac{\partial}{\partial \psi_2}\psi_1
=\omega\sum_{\alpha,\beta}\frac{\partial}{\partial\psi_\alpha}
(\tau_2)_{\alpha\beta}\psi_\beta.
\ee
Expressed in terms of the conjugate Grassmann variables, in a notation with two component spinors $\psi$, it reads
\be\label{P8A}
H=\omega\hat\psi ^T\tau_2\psi.
\ee

This can directly be used for the construction of the corresponding functional integral, with $\psi$ taken at $t+\epsilon$ and $\hat\psi $ at $t$ as in eq. \eqref{J6},
\be\label{156A}
H\big[\hat\psi (t),\psi(t+\epsilon)\big]=\omega\hat\psi ^T(t)\tau_2\psi(t+\epsilon).
\ee
We note that $H$ is now a bosonic quantity such that no time ordering $\hat T$ is needed for the functional integral \eqref{M1A}, with action
\ba\label{156B}
S=\sum_t\Big\{\hat\psi (t)^T\big(\psi(t+\epsilon)-\psi(t)\big)
+i\epsilon H\big[\hat\psi (t),\psi(t+\epsilon)\big]\Big\}.\nn\\
\ea
As it should be, the Hamiltonian \eqref{P8A} is hermitean
\be\label{158C}
\left(H[\hat\psi ,\psi]\right)^\dagger=H[\hat\psi ,\psi].
\ee

In the continuum limit we can write the action as a sum of a ``dynamical term'' $S_{dyn}$ and a ``Hamiltonian term'' $S_H$, 
\ba\label{158D}
S&=&S_{dyn}+S_H~,~S_{dyn}=\int dt\hat\psi (t)^T\partial_t\psi(t),\\
S_H&=&i\int dtH(t)~,~H(t)=H\big(\hat\psi (t),\psi(t)\big)=H^\dagger(t).\nn
\ea
This implies antihermiticity of $S=-S^\dagger$, and therefore hermiticity of the Minkowski action \eqref{117XB}. So far the action $S$ of our example is real and we therefore consider a real Grassmann algebra. 

\bigskip\noindent
{\bf 2. Complex structure}

\medskip\noindent
The time evolution \eqref{P3A} can also be described in a complex basis
\be\label{P9A}
\tilde q=
\left(\begin{array}{c}
q_1+iq_4\\q_2+iq_3
\end{array}\right)~,~
i\partial_t\tilde q=H\tilde q~,~H=\frac{\omega}{2}(\tau_3-1).
\ee
(Up to normalization, this corresponds to a spin in a constant magnetic field.) The definition of the complex two-component wave function $\tilde q$ in eq. \eqref{P9A} defines a complex structure for the four real components $q_\tau$. It is given by an involution $\theta$, whereby $q_3$ and $q_4$ are odd with respect to the involution (they change sign), whereas $q_1$ and $q_2$ are even (invariant). Not all possible observables for the real wave function $\{q_\tau \}$ are compatible with the complex structure. Compatibility means that observables can be expressed as complex operators $A$ acting on $\tilde q$, without involving the complex conjugate $\tilde q^*$. In other words, the expectation value of compatible observables obeys
\be \label{126A}
\langle A\rangle=\tilde q^\dagger A \tilde q.
\ee
An example for an observable which is compatible with the complex structure \eqref{P9A} is given in the real basis by $\{A_{\tau}\}=\{0,1,1,0\}$. In the quantum formalism this corresponds to the real diagonal operator 
\be\label{159B}
A=diag (0,1,1,0)=(N_1-N_2)^2.
\ee
It equals one if $N_1$ and $N_2$ are different, and zero if they are the same. In the complex basis it is represented as $A=(1-\tau_3)/2$. Since $H=-\omega A$ the quantum formalism implies that $\langle A\rangle$ is time independent. This can, of course, also be inferred directly from the time evolution of the probability distribution \eqref{P5A} and the values $A_{\tau}$ for the four states of the classical ensemble.

The complex structure for the wave function is reflected in a possible complex structure for the Grassmann algebra. The two-component wave function $(\tilde q_1, \tilde q_2)=(q_1+iq_4,q_2+iq_3)$  can now be associated to an element of a complex Grassmann algebra
\be \label{Y1A}
g=\tilde{q}_1+\tilde{q}_2\psi.
\ee
This algebra uses only one independent Grassmann variable $\psi$ instead of the two variables $\psi_1$, $\psi_2$ for the formulation in terms of a real Grassmann algebra. In the complex formulation the Schr\"odinger equation takes the form
\ba \label{Y2A}
i\partial_t g&=& {\cal H} g=\tilde{q}_2\psi, \\
{\cal H}&=& \omega\left(\frac{\partial}{\partial \psi}\psi-1 \right)=\omega({\cal N}-1).
\ea
In terms of the conjugate Grassmann variable this yields
\be \label{Y3A}
H=\omega(\hat\psi  \psi-1), \qquad H^{\dagger}=H,
\ee
and we can formulate the functional integral using
\be \label{Y4A}
S_H=i\omega\int dt (\hat\psi (t)\psi(t)-1).
\ee
All steps of sects. \ref{ConjugateGrassmannvariables} and \ref{quantumwavefunction} remain valid for this complex Grassmann algebra.

The equivalence of a complex wave function and Grassmann algebra with a real wave function and Grassmann algebra is easily generalized \cite{3A}. The real formulation has one Grassmann variable more than the complex one. Out of the $B$ bits in the real formulation one bit can be used for differentiating between the real and imaginary parts of a corresponding complex wave function. The remaining $B-1$ bits can be used for the construction of a basis of a $2^{B-1}$-component complex wave function. The latter is equivalent to a $2^B$-component real wave function. In consequence, real wave functions $\{q_\tau \}$ can cover the most general case of quantum wave functions. The complex wave functions and complex Grassmann algebra which are characteristic for many quantum systems reflect the presence of an ``additional'' complex structure.

We finally observe that an analytic continuation to ``euclidean time'' $\tau=it$ leaves $S_{dyn}$ invariant, while $S_H$ picks up a factor $-i$ from $dt$ (assuming $S_H$ contains no time derivatives)
\be\label{158E}
S_E=\int d\tau\big(\hat\psi \partial_\tau\psi+H(\hat\psi ,\psi)\big).
\ee
The corresponding euclidean functional integral realizes Osterwalder-Schrader positivity \cite{OS}. 

\section{Relating past and future}
\label{A16}
The formulation of a functional integral with one ``boundary condition'' $g_{in}$ set in the past, and the other one $\hat g_f$ set in the future needs the relation between $\hat g_f$ and $g_{in}$. If the measured observables and correlations involve only observables within a certain time interval, we can use $g_{in}(t_{in}),\hat g_f(t_f)$ with $t_{in}$ and $t_f$ at the boundaries of this interval. If we start from a functional integral which involves arbitrary times $(t_{in}\to-\infty; t_f\to\infty$), the wave functions $g_{in}(t_{in})$ and $\hat g_f(t_f)$ are the only pieces of information needed from the probability distribution outside the interval which is relevant for the observables \cite{3}. Still, for the most general case the relation between $\hat g_f$ and $g_{in}$ can be complicated, which would make a practical use of this formulation difficult.

\bigskip\noindent
{\bf 1. Time translations}

\medskip\noindent
For many relevant systems, however, the dynamics is characterized by a time independent Hamiltonian ${\cal H}$ or evolution operator ${\cal K}$. In the functional integral formalism this corresponds to an action that is invariant under time translations. In this case the factor $\hat g_f$ is easily computable for a large class of states specified by the initial wave function $g_{in}$.  If we assume the presence of a complex structure it is sufficient that $g_{in}$ is an eigenvalue of the Hamiltonian
\be\label{173A}
{\cal H}g_{in}=E g_{in}.
\ee
This implies the simple time evolution
\be\label{173B}
c_\tau(t)=e^{-iE(t-t_{in})}c_\tau(t_{in})
\ee
Thus $c_\tau(t_f),g(t_f)$ and $\hat g(t_f)$ are easily obtained from $g_{in}$ and the formulation of the functional integral \eqref{T10} can be employed without major obstacles in practice. A particularly simple example is the computation of vacuum properties, for which $g$ is time independent and therefore $\hat g_f=\hat g_{in}$. 

One also can choose $t_f-t_{in}=m\pi/E-\epsilon$ such that $g_f=\pm e^{i\epsilon E} g_{in},\hat g_f=\pm e^{-i\epsilon E} \hat g_{in}$, with the minus sign realized for $m$ odd. In this case one can put time on a torus, with periodic or antiperiodic boundary conditions on $g(t)$ for $m$ even or odd. Indeed, we can write the functional integral \eqref{T10} as
\be\label{182B}
\kl A(t)\kr=\int \cD\psi\cD\hat\psi A\big[\hat\psi(t),\psi(t)\big]e^{-S}\rho(t_{in}),
\ee
where we assume for simplicity that $A$ and $S$ involve only even powers of Grassmann variables. We identify $t_f=t_{in}-\epsilon$ such that the functional integral is periodic in time with period $m\pi/E$. We note that $\rho(t_{in})$ only involves $c_\tau(t_{in})$ but depends on the Grassmann variables $\hat\psi(t_f)=\hat\psi(t_{in}-\epsilon)$ and $\psi(t_{in})$. Furthermore, we want $S$ to be translation invariant around the torus such that we have to supplement the missing terms $\exp\Big\{-\hat\psi(t_{in}-\epsilon)\psi(t_{in})-i\epsilon H\big[\hat\psi(t_{in}-\epsilon),\psi(t_{in})\big]\Big\}$. We compensate this by the definition of $\rho(t_{in})$,
\ba\label{182C}
\rho(t_{in})&=&(-1)^m\sum_{\tau,\rho}c^*_\rho(t_{in})c_\tau(t_{in})\exp \big\{\hat\psi
(t_{in}-\epsilon)\psi(t_{in})\big\}\nn\\
&&\times \exp \Big\{i\epsilon\Big(H\big[\hat\psi(\tepi),\psi(t_{in})\big]-E\Big)\Big\}\nn\\
&&\times\hat g_\rho\big(\hat\psi(\tepi)\big)g_\tau\big (\psi(\ti )\big),
\ea
with $\hat g_\rho$ conjugate to $g_\rho$, such that 
\ba\label{182A}
\int d\psi(\ti )d\hat\psi(\tepi)\exp \big\{\hat\psi(\tepi)\psi(\ti )\big\}\nn\\
\times\hat g_\rho\big (\hat\psi(\tepi)\big)g_\tau\big 
(\psi(\ti )\big)=\delta_{\tau\rho},
\ea
and $\int d\psi(\ti )d\hat\psi(\tepi)\rho(\ti )=(-1)^m$. 

The expression \eqref{182B} computes $\kl A(t)\kr$ as a correlation function for the ``observables'' $A$ and $\rho(\ti )$ in a periodic functional integral. All information about the particular state is encoded in $\rho(\ti )$. We can easily generalize eq. \eqref{182B} to observables $A$ involving several time arguments $t_i$ inside the interval $\ti <t_i<\ti +m\pi/E$. This type of observables for energy eigenstates \eqref{173A} can be obtained with periodic time, and a ``long time behavior'' with $\Delta t> m\pi/E$ never matters. In analogy to the Matsubara formalism one may switch to frequency space with discrete frequencies $\omega_n=(2n+1)E/m,n\in {\mathbbm Z}$. We recall, however, that one works now  with ``real time'' without analytic continuation.

The presence of a complex structure is quite generic in our setting. Besides structures of the type discussed in the preceeding section, we can use charge conjugation as the involution underlying the complex structure. For our Ising type classical statistical systems charge conjugation maps empty bits to occupied bits and vice versa, $\tau=[n_\alpha]\to \bar\tau=[\bar n_\alpha]=\big[(1-n_\alpha)\big]$. On the level of the real Grassmann algebra we can define the charge conjugation as 
\be\label{182E}
\cC \psi_\alpha=\sigma_\alpha\tilde\psi_\alpha~,~
\sigma^2_\alpha=1~,~\cC^2=1.
\ee
This extends to arbitrary basis elements of the Grassmann algebra,
\be\label{182F}
\cC g_\tau=g^c_\tau=\sigma_\tau\tilde g_\tau,
\ee
for an appropriate choice of signs $\sigma_\tau=\pm 1$. If the evolution $\cK$ commutes with $\cC$ we can use $\cC$ for the definition of a complex structure - an example will be given in the next section.

\bigskip\noindent
{\bf 2. Time reversal}

\medskip\noindent
The elements $\hat g_f(t_f)$ and $g_{in}(\ti )$ may also be related by time reflection. In general, the dynamical term in the action is invariant under a time reflection around $t_s=(t_f+\ti )/2$,
\ba\label{182G}
T:&&\psi_\alpha(t_s+t)\to \sum_\beta T_{\alpha\beta}\hat\psi_\beta(t_s-t),\nn\\
&&\hat\psi_\alpha(t_s+t)\to \sum_\beta\hat T_{\alpha\beta}\psi_\beta(t_s-t),
\ea
where $\hat T$ and $T$ are related by $\hat T^T=T^{-1}$, accompanied by a total reordering of all Grassmann variables. The time reflection is an involution, $T^2=1$. If for an appropriate choice of $T$ and $\hat T$ also the Hamiltonian part $S_H$ remains invariant, the action is time reflection invariant $T(S)=T$. For the action \eqref{101B}  this is realized by $T=\hat T=-1$, while for the action \eqref{156B} we use in eq. \eqref{182G} the matrices
\be\label{182H}
T=\hat T=\tau_1.
\ee
(Our definitions are such that reflection symmetry holds in the formulation with discrete time steps $t_n$ without further modifications.) The functional integral is $T$-invariant if
\be\label{182I}
T\big (\hat g_f(t_f)\big)=\pm g_{in}(\ti ) ~,~T\big (g_{in}(\ti )\big)=\pm \hat g_f(t_f).
\ee
We emphasize that time reflection symmetry is a property of the whole family of probability distributions $\big \{p_\tau(t)\big\}$ for all $t$, rather than being a symmetry of a state at a given $t$. 

For a time reflection invariant action the condition  \eqref{182I} is obeyed if
\be\label{182J}
T\big(g(t_s)\big)=\pm\hat g(t_s).
\ee
This follows from the reflection symmetry of the time evolution. The easiest way to see that eq. \eqref{182J} follows from eq. \eqref{182I} relies on the definition \eqref{73B}, realizing that $S_<$ is mapped by time reflection to $S_>,T(e^{-S_<})=e^{-S_>}$. The inverse follows from the invertibility of the time evolution. For our first example and $t_s=0$ a time reflection symmetric setting is realized for $g(0)=1$ or $g(0)=\psi$, corresponding to $p_1(0)=1$ or $p_0(0)=1$. One verifies that the corresponding time evolution of the probability distribution \eqref{61} is indeed reflection symmetric. For the second example in eq. \ref{Twocomponentspinor} we have reflection symmetry for $g(0)=\frac{1}{\sqrt{2}}(\psi_1\pm \psi_2)$, corresponding to $p_2(t)=p_3(t)$ in eq. \eqref{P5A}. If eq. \eqref{182J} is obeyed for some suitable $t_s$ we may choose $t_f$ and $\ti $ obeying $t_s=(t_f+\ti )/2$. Then eq. \eqref{182I} allows us to compute $\hat g_f(t_f)$ in terms of $g_{in}(\ti )$ and the functional integral is fully specified by the coefficients $c_\tau(\ti )$. We observe in this context that the choice of the matrices $T$ and $\hat T$ in eq. \eqref{182G}, which leave $S$ reflection invariant, may not be unique. Any choice which realizes eq. \eqref{182J} will do. 

\section{Quantum field theory for two-dimensional fermions}
\label{qft}
In this section we generalize our previous examples to a quantum field theory for free fermions in two dimensions (one space and one time dimension). We will construct a map between the functional integral defining a quantum  field theory for fermions and an associated probability distribution of a classical statistical ensemble for Ising type variables. Further generalizations to fermionic quantum field theories with interactions and in arbitrary dimensions are straightforward. We use a real Grassmann algebra and the index $\alpha$ covers a space index $x_i$ on a one-dimensional lattice with $x_{i+1}-x_i=\epsilon$, as well as an internal index $\gamma=\pm$. We may use a torus with $L$ points $x_i$ and circumference $l=L\epsilon$, such that $B=2L$. In the continuum limit we deal with two Grassmann valued fields $\psi_+(t,x),\psi_-(t,x)$. The functional integral involves only real elements, even though we use $i$ occasionally for the purpose of analogy with quantum mechanics.

\bigskip\noindent
{\bf 1. Functional integral and Schr\"odinger equation}

\medskip\noindent
In a discrete setting the action in the functional integral formulation is given by
\be \label{D1}
\begin{split}
S= \sum_{t,x} &\big\{\hat{\psi}_+(t,x)\left(\psi_+(t+\epsilon,x-\epsilon)-\psi_+(t,x)\right) \\
&+\hat{\psi}_-(t,x-\epsilon)\left(\psi_-(t+\epsilon,x)-\psi_-(t,x-\epsilon)\right) \big\}
\end{split}
\ee
or equivalently
\be \label{D1A}
\begin{split}
S= \sum_{t,x} &\big\{\hat{\psi}_+(t,x)\left(\psi_+(t+\epsilon,x)-\psi_+(t,x)\right) \\
&+\hat{\psi}_-(t,x-\epsilon)\left(\psi_-(t+\epsilon,x-\epsilon)-\psi_-(t,x-\epsilon)\right) \\
&-\hat{\psi}_+(t,x)\left(\psi_+(t+\epsilon,x)-\psi_+(t+\epsilon,x-\epsilon)\right) \\
&+\hat{\psi}_-(t,x-\epsilon)\left(\psi_-(t+\epsilon,x)-\psi_-(t+\epsilon,x-\epsilon)\right) \big\}.
\end{split}
\ee
The continuum limit involves now a rescaling of the Grassmann variables $\psi_\pm(t,x)\to \sqrt{\epsilon}\psi_\pm(t,x)$ such that
\ba \label{D2}
S&=& \int_{t,x} \big\{\hat\psi _+\partial_t\psi_+ +\hat\psi _-\partial_t\psi_- -\hat\psi _+\partial_x\psi_+ +\hat\psi _-\partial_x\psi_- \big\} \nn \\
&=&\int_{t,x}\psi^{\dagger}\partial_t \psi +i \int_t H,
\ea
with $\psi=(\psi_+,\psi_-)$, $\psi^{\dagger}=(\hat\psi _+,\hat\psi _-)$. The Hamiltonian part reads
\be \label{D3}
H=i\int_x \big\{\hat\psi _+\partial_x \psi_+-\hat\psi _-\partial_x \psi_- \big\}=i\int_x \psi^\dagger \tau_3\partial_x \psi,
\ee 
and obeys $H^\dagger=H$, $H^*=H^{T}=-H$, such that $S^T=-S$ and $S^\dagger_M=S_M$. The functional integral is defined as in the preceding sections. 

The Schr\"odinger equation for the Grassmann wave function $g$ obtains as
\ba \label{D4}
\partial_t g&=&-i {\cal H}g, \nn \\
{\cal H}&=& i\int_x \big\{\frac{\partial}{\partial \psi_+}\partial_x \psi_+-\frac{\partial}{\partial \psi_-}\partial_x \psi_- \big\},
\ea
where we replace $\hat\psi _\pm$ in eq. \eqref{D3} by $\partial/\partial\psi_\pm$. The Hamilton operator ${\cal H}$ commutes with the two ``particle numbers''
\be\label{D4a}
\bar{\cal N}_+=\int_x \frac{\partial}{\partial \psi_+(x)} \psi_+(x), \qquad \bar{\cal N}_-=\int_x \frac{\partial}{\partial \psi_-(x)} \psi_-(x), \nonumber
\ee
\be \label{D5}
[{\cal H}, \bar{\cal N}_\pm]=0.
\ee
Thus $\bar{\cal N}_+$ and $\bar{\cal N}_-$ are conserved. We can consider simultaneous eigenvectors of $\bar{\cal N}_+$ and $\bar{\cal N}_-$
\be \label{D6}
\bar{\cal N}_+ g_{n_+,n_-}=n_+ g_{n_+,n_-}, \qquad \bar{\cal N}_- g_{n_+,n_-}=n_- g_{n_+,n_-}  
\ee
and expand a general wave function as
\be \label{D7}
g=\sum_{n_+, n_-}A_{n_+,n_-}g_{n_+,n_-}.
\ee

\bigskip\noindent
{\bf 2. Multiparticle states}

\medskip\noindent
The one particle state with $n_+=1$, $n_-=0$ can be written as
\be \label{D8}
g_{1,0}=\int_x q_{1,0}(x)\frac{\partial}{\partial \psi_+(x)}|0\rangle=\int_x q_{1,0}(x)a^\dagger_+(x)|0\rangle
\ee
with vacuum state
\be \label{D9}
|0\rangle=\prod_x \psi_+(x)\psi_-(x).
\ee
Here we also employ the notation of sect. \ref{fermions} by associating $\partial/\partial \psi$ with the creation operator $a^{\dagger}$. The real one particle wave function is given by $q_{1,0}(t,x)$ and normalized according to
\be \label{D8A} 
\int_x q_{1,0}^2(x)=1. 
\ee 
On the torus we use antiperiodic boundary conditions for the wave functions as $q_{1,0}(x)$, such that the associated probabilities as $p_{1,0}(x)$ are periodic.

The wave function $q_{1,0}$ obeys a one-particle Schr\"odinger equation which is derived by inserting $g=g_{1,0}$ in eq. \eqref{D4}. This becomes a very simple differential equation
\be \label{D10}
\partial_t q_{1,0}(t,x)=\partial_x q_{1,0}(t,x).
\ee
The general solution $q_{1,0}(x+t)$ depends only on $x+t$ and therefore describes a ``left-moving'' particle. The arbitrary shape of the wave function is conserved and specified at some initial time $t_0$ by $q_{1,0}(t_0,x)$. Similarly, $g_{0,1}$ describes ``right-moving'' particles for which the real wave function $q_{0,1}(x-t)$, which solves the Schr\"odinger equation
\be \label{D11}
\partial_t q_{0,1}=-\partial_x q_{0,1},
\ee
depends only on $x-t$.

Two left-moving particles involve the two-particle wave function $q_{2,0}(x,y)$,
\be \label{D12}
g_{2,0}=\frac{1}{\sqrt{2}}\int_{x,y} q_{2,0}(x,y)\frac{\partial}{\partial \psi_+(x)}\frac{\partial}{\partial \psi_+(y)}|0\rangle,
\ee
with normalization
\be \label{D12A}
\int_{x,y}q_{2,0}^{2}(x,y)=1.
\ee
As appropriate for two identical fermions it is antisymmetric under particle exchange
\be \label{D13}
q_{2,0}(x,y)=-q_{2,0}(y,x).
\ee
In contrast, one left-moving particle at $x$ and one right-moving particle at $y$ differs from the state where the left-mover is at $y$ and the right-mover at $x$. Thus for
\be \label{D14}
g_{1,1}=\int_{x,y}q_{1,1}(x,y)\frac{\partial}{\partial \psi_+(x)}\frac{\partial}{\partial \psi_-(y)}|0\rangle
\ee
the wave functions $q_{1,1}(x,y)$ and $q_{1,1}(y,x)$ are not related by symmetry. We encounter the standard situation for different particle species.  

The Schr\"odinger equation for two left-movers becomes
\be \label{D15}
\partial_t q_{2,0}(t,x,y)=(\partial_x+\partial_y)q_{2,0}(t,x,y).
\ee
A special solution is given by a factorized wave function
\be \label{D16}
q_{2,0}(t,x,y)=a(x+t)b(y+t)-a(y+t)b(x+t).
\ee
More general solutions obtain by superposition of terms of the type \eqref{D16}. The generalization to states with $n_+$ left-movers and $n_-$ right-movers is straightforward. Our normalizations are chosen such that the general Grassmann element \eqref{D7} obeys
\be \label{D16A}
\begin{split}
&\sum_{n_+,n_-}A^2_{n_+,n_-}=1, \\
&\int {\cal D}\psi \tilde g_{n'_+,n'_-} g_{n_+,n_-}=\delta_{n_+,n'_+}\delta_{n_-,n'_-}.
\end{split}
\ee
It is obvious that $q_{n_+,0}$ is the wave function for $n_+$ identical fermions, with total antisymmetry with respect to the exchange of any two particles. 

\bigskip\noindent
{\bf 3. Classical wave function and probability 

~distribution}

\medskip\noindent
We next turn to the classical statistical ensemble which is described by the action \eqref{D1}, with probability distribution $\{p_\tau\}$ and associated classical wave function $\{q_\tau\}$. The general wave function $q_\tau(t)$ evolves according to the Schr\"odinger equation obtained from eq. \eqref{D4} by inserting
\be \label{D17}
g(t)=\sum_\tau q_\tau(t)g_\tau
\ee
The states $\tau$ of the classical statistical ensemble are given by the sequence of occupation numbers, $\tau=[n_+(x),n_-(x)]$. The particle numbers $n_+$, $n_-$ in a given state $\tau$ obey
\be \label{D18}
n_+=\sum_x n_+(x), \qquad n_-=\sum_x n_-(x). 
\ee
For any given time $t$ we decompose 
\be \label{D19}
g(t)=\sum_{n_+,n_-}A_{n_+,n_-}(t)g_{n_+,n_-}(t),
\ee
where $g_{n_+,n_-}$ is normalized such that $p_{n_+,n_-}=A^2_{n_+,n_-}$ equals the probability to find $n_+$ left-moving particles and $n_-$ right-moving particles. Since the particle numbers $N_+$, $N_-$ are conserved, the solutions of the Schr\"odinger equation have time-independent amplitudes $A_{n_+, n_-}$. In turn, the general solution for the fixed particle wave function $q_{n_+, n_-}$, defined by
\ba\label{157A}
&&g_{n_+,n_-}(t)= (n_+! n_-!)^{-\frac{1}{2}}\nn \\ 
&&\times \int_{x_1\dots x_{n_+},y_1\dots y_{n_-}}q_{n_+,n_-}
(x_1\dots x_{n_+},y_1\dots y_{n_-},t) \nn\\
&&\times g_{n_+,n_-}(x_1\dots y_{n_-})
\ea
with
\ba\label{D20}
&&g_{n_+,n_-}(x_1\dots y_{n_-})=\\
&&\frac{\partial}{\partial \psi_+(x_1)}\dots\frac{\partial}{\partial \psi_+(x_{n_+})}\frac{\partial}{\partial \psi_-(y_1)}\dots\frac{\partial}{\partial \psi_-(y_{n_-})}|0\rangle\nn
\ea
have been discussed above. The coefficients $A_{n_+,n_-}$ and the wave functions $q_{n_+,n_-}(t)$ fix completely the evolution of the probability distribution $p_\tau(t)$.

In the opposite direction we may start with a given solution $p_\tau(t)$ of the evolution equation. From this probability distribution we can compute directly the conserved probabilities $p_{n_+,n_-}$ by summing the probabilities of all states with a given number $n_+$ of occupied bits of type $+$ and similar for the type $-$. Without loss of generality we can choose a convention where $A_{n_+,n_-}=\sqrt{p_{n_+,n_-}}\ge 0$. From the relative probabilities to find the $n_+$ occupied bits at positions $x_1\dots x_{n_+}$ and the $n_-$ occupied bits at $y_1\dots y_{n_-}$ we can determine the wave function $q_{n_+,n_-}(x_1\dots y_{n_-})$ up to a sign. The overall signs of the wave functions $q_{n_+,n_-}(t_0)$ at some initial time $t_0$ are free. The expectation values and correlations of observables do not depend on these signs and we may choose them according to some arbitrary convention. (This is similar to a choice of gauge). For all times $t$ the sign of $q_{n_+,n_-}(t)$ is then related to $q_{n_+,n_-}(t_0)$ by the solution of the Schr\"odinger equation. Furthermore, at $t_0$ only the overall sign of $q_{n_+,n_-}(t_0)$ is arbitrary. The relative signs for different arguments $(x_1,\dots x_{n_+},y_1,\dots y_{n_-})$ can be fixed by requirements of continuity and differentiability \cite{CWP}. Up to some irrelevant ``choice of gauge'' the wave function $\{q_{\tau}(t) \}$ for a system of free particles in two dimensions is determined by the probability distribution $\{p_\tau(t)\}$ of a classical statistical ensemble. For our example, the real quantum wave function $\psi_Q(t)=\sum_{n_+,n_-}A_{n_+,n_-}q_{n_+,n_-}(t)$ for a multi-fermion system is directly related to the classical wave function $\{q_{\tau} \}$. If the time evolution of the probability distribution $\{p_{\tau} \}$ is such that the associated Grassmann element \eqref{D19} obeys eq. \eqref{D4}, we can express all relevant expectation values by a functional integral with action \eqref{D1}.

\bigskip\noindent
{\bf 4. Lorentz symmetry}

\medskip\noindent
Our system of free fermions is invariant with respect to two-dimensional Lorentz transformations. This is most easily seen by using
\be \label{D21}
\bar\psi=(-\hat\psi _-,\hat\psi _+)=\psi^\dagger \gamma^0.
\ee
In terms of the real Dirac matrices
\be \label{D22}
\gamma^0=i\tau_2, \qquad \gamma_1=\tau_1,
\ee
which obey
\be \label{D23}
\{ \gamma^\mu,\gamma^\nu \}=2\eta^{\mu\nu}
\ee
with $\eta^{\mu\nu}=diag(-1,1)$ we can write
\be \label{D24}
S=-\int_{t,x}\bar\psi \gamma^\mu\partial_\mu\psi.
\ee
The infinitesimal Lorentz-transformations act on the real two component spinors $\psi$ and $\bar\psi$ as
\be \label{D25}
\delta\psi=-\epsilon \Sigma^{01}\psi, \qquad \delta\bar\psi=\epsilon \bar \psi \Sigma^{01},
\ee
with
\be \label{D26}
\Sigma^{01}=-\frac{1}{4}\big[ \gamma^0,\gamma^1\big]=-\frac{1}{2}\tau_3.
\ee
They commute with $\bar \gamma$ obeying
\be \label{D27}
\bar\gamma=\gamma^0\gamma^1=\tau_3, \quad \bar\gamma^2=1, \quad \{\gamma^\mu,\bar{\gamma}\}=0.
\ee
The eigenstates of $\bar\gamma$ correspond therefore to irreducible representations of the Lorentz group
\be \label{D28}
\begin{split}
& \left(\begin{array}{l}\psi_+\\0\end{array}\right)\,=\,\frac{1}{2}(1+\bar{\gamma})\psi, \qquad \left(\begin{array}{l}0\\ \psi_-\end{array}\right)\,=\,\frac{1}{2}(1-\bar{\gamma})\psi, \\
& \left(\begin{array}{l}0,\hat\psi _+\end{array}\right)\,=\,\frac{1}{2}\bar\psi(1-\bar{\gamma}), \qquad \left(\begin{array}{l} -\hat\psi _-, 0\end{array}\right)\,=\,\frac{1}{2}\bar\psi(1+\bar{\gamma}).
\end{split}
\ee
The Weyl-spinors $\psi_\pm$ transform by multiplicative rescalings
\be \label{D29}
\begin{split}
& \delta\psi_+=\frac{\epsilon}{2}\psi_+, \qquad \delta\psi_-=-\frac{\epsilon}{2}\psi_-, \\
& \delta\hat\psi _+=\frac{\epsilon}{2}\hat\psi _+, \qquad \delta\hat\psi _-=-\frac{\epsilon}{2}\hat\psi _-.
\end{split}
\ee
In addition, the coordinates $x^\mu=(t,x)$ are transformed in the usual way.

\bigskip\noindent
{\bf 5. Discrete symmetries}

\medskip\noindent
Besides the invariance under the continuous Lorentz-transformations the action \eqref{D2}, \eqref{D24} is also invariant under the discrete symmetries of parity reflection ($P$), time reversal ($T$) and charge conjugation ($C$). Parity reflection acts by reversing the sign of the space coordinate, $x\to -x$, accompanied by a discrete map acting on the Grassmann variables
\be \label{D30}
\begin{split}
& P\left(\psi(t,x)\right)=-\gamma^0\bar{\gamma}\psi(t,-x)=\tau_1\psi(t,-x), \\
& P\left(\bar{\psi}(t,x)\right)=-\bar{\psi}(t,-x)\bar{\gamma}\gamma^0=-\bar{\psi}(t,-x)\tau_1.
\end{split}
\ee
It exchanges two Weyl spinors $\psi_+$, $\psi_-$,
\be \label{D31}
P: \quad \psi_+ \leftrightarrow \psi_-, \qquad \hat{\psi}_+\leftrightarrow\hat{\psi}_-.
\ee
Time reversal changes the sign of $t$ and maps
\be \label{D32}
T: \quad  \psi_+\leftrightarrow \hat{\psi}_-, \qquad \psi_-\leftrightarrow \hat{\psi}_+.
\ee
In addition, all Grassmann variables are totally reordered (transposition).

Charge conjugation exchanges the Grassmann variables with their conjugate ones (no action on the coordinates)
\be \label{D33}
C: \quad \psi_+\leftrightarrow \hat{\psi}_+, \qquad \psi_-\leftrightarrow \hat{\psi}_-.
\ee
(This corresponds to a charge conjugation matrix $C_1=\gamma^0$ in the general setting of ref. \cite{CWES}. The use of the same name of charge conjugation should not hide the fact that the transformation \eqref{D33} is conceptually different from the charge conjugation discussed in sects. \ref{Symmetries} and \ref{A16}. It acts within the extended Grassmann algebra constructed from $\psi$ and $\hat\psi$, in contrast to eq. \eqref{S8a}.)

The combined symmetry $PTC$ maps $\psi_\gamma(t,x)\to \psi_\gamma(-t,-x),\hat\psi_\gamma(t,x)\to\hat\psi_\gamma(-t,-x)$, together with a transposition. Furthermore, the action \eqref{D2} is invariant under the reflections
\be \label{D35A}
\begin{split}
& S_+: \quad \psi_+\to-\psi_+, \qquad \hat{\psi}_+\to-\hat\psi _+, \\
& S_-: \quad \psi_-\to-\psi_-, \qquad \hat{\psi}_-\to-\hat\psi _-.
\end{split}
\ee
This allows us to define modified discrete transformations as $\tilde{P}_-=S_-P$ etc. Of course, the action remains invariant under these modified transformations as well. Finally, the transformation $\psi\to\gamma\psi,\bar\psi\to\bar\psi\gamma$, combined with a transposition, leaves $S$ invariant. This may be used for defining a version of time reflection without transposition. 

A simple extension of our model could use a complex Grassmann algebra with a complex wave function $\{c_\tau \}$ instead of the real $\{q_\tau\}$. For free fermions the real and imaginary parts of $\{c_\tau \}$ evolve independently and the extension accounts for a trivial doubling of the degrees of freedom. We could now add a pointlike interaction, $S\to S+S_I$,
\be \label{D36}
\begin{split}
S_I&=i\lambda\sum_{t,x}\hat{\psi}_+(t,x)\hat{\psi}_-(t,x)\psi_-(t+\epsilon,x)\psi_+(t+\epsilon,x) \\
&=i\lambda \int_{t,x}\hat{\psi}_+\hat{\psi}_-\psi_-\psi_+.
\end{split}
\ee
This interaction is invariant under Lorentz-transformations as well as the discrete transformations $P$, $T$ and $C$. The interaction part of the Hamiltonian
\be \label{D37}
H=\lambda\int_x \hat{\psi}_+\hat{\psi}_-\psi_-\psi_+
\ee
is hermitean only for real $\lambda$. This is the reason why a pointlike interaction is not possible for a real Grassmann algebra. In the following we will not discuss this extension. We will rather discuss a complex structure which allows to represent the real wave function $\{q_\tau\}$ in terms of a complex wave function $\{c_\tau\}$ with half the number of components.

\bigskip\noindent
{\bf 6. Weyl spinors}

\medskip\noindent
We observe that the action for free fermions can be written as a sum
\be \label{D38}
S=S_++S_-
\ee
where $S_+$ only involves the left-movers $\psi_+$ and $\hat{\psi}_+$, whereas $S_-$ accounts for the right-movers $\psi_-$ and $\hat{\psi}_-$. The functional measure can be written as a product of integrals, one involving $\psi_+$, $\hat{\psi}_+$ and the other $\psi_-$, $\hat{\psi}_-$. If the initial state factor $g_{in}$ has a factorized form
\be \label{D39}
g_{in}=g\left( \psi(t_{in})\right)=g_+\left(\psi_+(t_{in}) \right)g_-\left(\psi_-(t_{in})\right)
\ee
this also holds for $\hat{g}_f=\hat{g}\left( \hat{\psi}(t_f)\right)$ and the partition function \eqref{M1A} factorizes
\be \label{D40} 
Z=Z_+Z_-.
\ee
In this case our setting describes independent ``worlds'' for the left-movers and right-movers. The wave function factorizes
\be \label{D41}
q_\tau(t)=q_\rho^{(+)}(t)q_\sigma^{(-)}(t), \qquad \tau=(\rho,\sigma)
\ee
with $\rho=[n_+(x_i)]$, $\sigma=[n_-(x_i)]$. This extends to the probability density
\be \label{D42}
p_\tau(t)=p_\rho^{(+)}(t)p_\sigma^{(-)}(t), \qquad p_\rho^{(\pm)}(t)=\left(q_\rho^{(\pm)} \right)^2.
\ee

In fact, a model involving only the right-moving Weyl spinors $\psi_-$, $\hat\psi _-$ is fully consistent. It obtains from our setting by removing $\psi_+$ and $\hat{\psi}_+$, i.e. by putting $\psi_+=\hat\psi _+=0$ in the action and omitting $\psi_+$, $\hat{\psi}_+$ in the functional measure. The action is then invariant under Lorentz-transformations and charge conjugation, but violates $P$ and $T$ since these symmetries map the right-movers to the left-movers. A model involving only the Weyl spinors $\psi_-$, $\hat{\psi}_-$ still conserves the discrete symmetry
\be \label{D43}
PT: \quad \psi_-\leftrightarrow \hat{\psi}_-, 
\ee
and is therefore also invariant under $PTC$.

In two-dimensions Majorana-Weyl spinors are compatible with Lorentz-transformations \cite{CWMS}. They obey $C\psi=\psi$ and obtain by identifying $\hat{\psi}_+$ and $\psi_+$ as well as $\hat{\psi}_-$ and $\psi_-$. It is not yet clear what would be the probability interpretation for an action based on such Majorana-Weyl spinors.

\bigskip\noindent
{\bf 7. Antiparticles}

\medskip\noindent
For every classical state $\tau=[n_+(x),n_-(x)]$ we can define the ``antistate'' $\bar\tau=[\bar n_+(x),\bar n_-(x)]$ with $\bar n_\pm =1-n_\pm$. The mapping $\tau\to\bar\tau$ transforms occupied bits into empty bits and vice versa. In the language of fermionic excitations it exchanges particles and holes or particles and antiparticles. A state with particle numbers $(n_+,n_-)$ is mapped into a state with $\bar n_+=n_+$ antiparticles of type $+$ and $\bar n_-=n_-$ antiparticles of type $-$. For $L$ space points $x_i$ a state with $\bar n_+$ antiparticles has $L-\bar n_+$ particles of type $+$, and similar for $\bar n_-$. For every state the number of particles plus holes for each species equals $L$, $n_++\bar n_+=L~,~n_-+\bar n_-=L$, since for every space point there is either an empty or an occupied bit.

The state with one antiparticle of type $+$, i.e. $(\bar n_+,\bar n_-)=(1,0)$, can be written in the Grassmann formulation as
\be\label{S1}
g^c_{1,0}=\int_xq^c_{1,0}(x)g^c_{1,0}(x),
\ee
with $g^c_{1,0}=\tilde g_{1,0}$ the Grassmann element conjugate to $g_{1,0}$, obeying eq. \eqref{26}, 
\be\label{S2}
\int{\cal D}\psi\tilde g_{1,0}(x)g_{1,0}(y)=\delta(x-y).
\ee
One finds the simple expression
\be\label{S3}
g^c_{1,0}(x)=\psi_+(x).
\ee
Reordering the Hamiltonian \eqref{D3}, 
\be\label{S4}
{\cal H}=-i\int_x\left\{\partial_x\psi_+(x)
\frac{\partial}{\partial\psi_+}(x)-\partial_x\psi_-(x)
\frac{\partial}{\partial\psi_-}(x)\right\},
\ee
the wave function for one antiparticle obeys the same evolution law as for a particle,
\be\label{S5}
\partial_t q^c_{1,0}=\partial_x q^c_{1,0}.
\ee
Both the particle and the hole of type $+$ are left-movers. Similarly, the wave function $q^c_{0,1}$ for one antiparticle of type $-$ obeys $\partial_tq^c_{0,1}=-\partial_x q^c_{0,1}$ such that both the particle and the antiparticle of type $-$ are right-movers. 

More generally, a state with $\bar n_+$ antiparticles of type $+$ and $n_-$ antiparticles of type $n_-$ can be written in analogy with eq. \eqref{157A} as
\ba\label{S6}
g^c_{\bar n_+, \bar n_-}&=&
(\bar n_+!\bar n_-!)^{-1/2}
\int_{x_1\dots x_{\bar n_+},y_1\dots y_{\bar n_-}}\nn\\
&&q^c_{\bar n_+,\bar n_-}(x_1\dots x_{\bar n_+},y_1\dots y_{\bar n_-})\nn\\
&&g^c_{\bar n_+,\bar n_-}(x_1\dots x_{\bar n_+},y_1\dots y_{\bar n_-}),
\ea
with
\ba\label{S7}
&&g^c_{\bar n_+,\bar n_-}(x_1\dots x_{\bar n_+},y_1\dots y_{\bar n_-})=
s_c\tilde g_{\bar n_+,\bar n_-}(x_1\dots y_{\bar n_-})\nn\\
&&=s_c\psi_-(y_{\bar n_-})\dots 
{\psi}_-(y_1)\psi_+(x_{\bar n_+})\dots \psi_+(x_1),
\ea
whereby $s_c=1$ for $\bar n_++\bar n_-=1,4$ mod $4$ and $s_c=-1$ for $\bar n_++\bar n_-=2,3$ mod $4$. For $\bar n_+=n_+,\bar n_-=n_-$ the time evolution of $q^c_{\bar n_+,\bar n_-}$ and $q_{n_+,n_-}$ are given by identical (real) evolution equations, 
\ba\label{S8}
&&\partial_t q^c_{\bar n_+,\bar n_-}(x_1\dots y_{n_-})=
K_{n_+,n_-}q^c_{\bar n_+,\bar n_-}(x_1\dots y_{n_-})\nn\\
&&\partial_t q_{n_+,n_-}(x_1\dots y_{n_-})=
K_{n_+,n_-}q_{n_+,n_-}(x_1\dots y_{n_-}).\nn\\
\ea

We define the operation of the charge conjugation $C$ on the level of the Grassmann algebra $g[\psi]$ (without $\hat\psi$) as
\be\label{S9}
{\cal C}g_{n_+,n_-}(x_1\dots y_{n_-})=g^c_{n_+,n_-}(x_1\dots y_{n_-})~,~{\cal C}^2=1,
\ee
implementing a mapping between particle and associated antiparticle states. The Hamiltonian commutes with ${\cal C}$,
\be\label{S10}
[{\cal H},{\cal C}]=0~,~[K_{n_+,n_-},C]=0,
\ee
where we observe that for any Grassmann operator 
${\cal A}[a,a^\dagger]\equiv{\cal A}\left[\psi,\frac{\partial}{\partial \psi}\right]$ the charge conjugate operator ${\cal A}^c={\cal C}{\cal A}{\cal C}$ obtains by exchanging $a$ and $a^\dagger$ or $\psi$ and $\partial/\partial\psi$ (keeping the order). For example, in a discrete formulation the particle number operators \eqref{D4a} transform as 
\be\label{S11}
\bar {\cal N}^c_\pm={\cal C}\bar{\cal N}_\pm{\cal C}=L-\bar{\cal N}_\pm.
\ee
The charge conjugate of the vacuum state $|0\kr$ \eqref{D9} is the unit element $|1\kr=1$ of the Grassmann algebra, ${\cal C}|0\kr=|1\kr,{\cal C}|1\kr=|0\kr$, and $g^c_{\bar n_+,\bar n_-}$ obtains as
\ba\label{S12}
&&g^c_{\bar n_+,\bar n_-}(x_1\dots y_{\bar n_-})=
{\cal C}g_{\bar n_+,\bar n_-}(x_1\dots y_{\bar n_-})\nn\\
&&~\quad={\cal C}\frac{\partial}{\partial\psi_+(x_1)}{\cal C}\dots{\cal C}
\frac{\partial}{\partial\psi_-(y_{\bar n_-})}
{\cal C}{\cal C}|0\kr\nn\\
&&~\quad=\psi_+(x_1)\dots\psi_-(y_{\bar n_-}).
\ea
On the level of the wave functions the charge conjugation maps 
$q_\tau\to q^c_\tau=(Cq)_\tau$ and we can formally write
\be\label{S13}
g^c=\sum_\tau q^c_\tau g_\tau=\sum_\tau(C q)_\tau g_\tau
=\sum_\tau q_\tau g^c_\tau.
\ee

\bigskip\noindent
{\bf 8. Complex structure}

\medskip\noindent
The evolution equation \eqref{S8} consists of two real equations for the real functions $q$ and $q^c$, which have the same arguments. We can combine them into one complex equation. For this purpose we define the complex wave function
\be\label{S14}
c\n=\frac12 \left\{
q\n+q^c\n+i(q\n-q^c\n)\right\}.
\ee
The real part of $c$ is even under charge conjugation, while the imaginary part is odd. In this complex basis we can represent the charge conjugation as complex conjugation
\be\label{S15}
Cc\n=c^*\n.
\ee
The evolution equation \eqref{S8}  becomes now a complex equation 
\be\label{S16}
\partial_tc\n(x_1\dots y_{n_-})=Kc\n(x_1\dots y_{n_-}).
\ee
The normalization of $c\n$ is chosen such that
\be\label{S17}
\int_{x_1\dots y_{n_-}}|c\n(x_1\dots y_{n_-})|^2=1.
\ee

More generally, a complex structure within a Grassmann algebra is defined by an involution $\theta ,\theta^2=1$, which maps an arbitrary element of the Grassmann algebra $g$ into another one, $\theta(g)$. In our case the involution is defined by the charge conjugation, 
$\theta(g)=g^c=\sum_\tau q_\tau g^c_\tau=\sum_\tau q^c_\tau g_\tau$. The presence of $\theta$ allows a map from a real Grassmann algebra to a complex Grassmann algebra, $q_\tau\to c_\tau$. The elements which are even with respect to $\theta$ are mapped to $Re(c_\tau)$, while $\theta$-odd elements are mapped to the imaginary part $Im(c_\tau)$,
\be\label{S18}
Re(c_\tau)=\frac12(q_\tau+q^c_\tau)~,~Im(c_\tau)=\frac12(q_\tau-q^c_\tau).
\ee

For the complex representation of the Grassmann algebra we only need one half of the basis elements $g_\tau$. Indeed, from the knowledge of the complex coefficients $c\n$ we can extract both the real coefficients $q\n$ and $q^c\n=\pm q_{L-n_+,L-n_-}$. We therefore need only the basis elements $g\n$ with $n_++n_-\leq L$, since the associated $c\n$ also specifies $q\n$ with $L\leq n_++n_-\leq 2L$. For $n_++n_-=L$ we retain as independent basis elements the ones with $n_+\geq L/2$. For the special case $n_+=n_-=L/2$ - for even $L$ - the charge conjugation flips between empty and filled positions and we only need half of them as independent elements. With this restriction to independent Grassmann elements the complex Grassmann algebra (with complex wave function $c_\tau$) and the real Grassmann algebra (with real wave function $q_\tau$) are equivalent. Each real Grassmann algebra admits the charge conjugation as a natural possible complex structure. 

In the complex formulation we can exploit the structure of multiplication with complex numbers or the complex Fourier transform. In particular, the multiplication with $i$ corresponds in the real formulation to a map $q_\tau\to q^c_\tau,q^c_\tau\to -q_\tau$,
\ba\label{S19}
&&c_\tau\to ic_\tau~\widehat{=}~
\left(\begin{array}{c}q_\tau\\q^c_\tau\end{array}\right)
\to I
\left(\begin{array}{c}q_\tau\\q^c_\tau\end{array}\right),\nn\\
&&I=\left(\begin{array}{cc}0&1\\-1&0\end{array}\right)~,~I^2=-1.
\ea
This is not equivalent to the formal multiplication with $i$ in the real formulation of the Grassmann algebra. The latter has been employed above in order to illustrate the analogy to the quantum formalism even though all equations remain real equations. (Only the real operator $K=-iH$ appears in the evolution equations.) In contrast, whenever we work with a complex Grassmann algebra the multiplication with $i$ is meant in the sense of eq. \eqref{S19}. Using the complex formulation we recover the standard complex wave functions for multi-fermion states.

\section{Conclusions}
\label{conclusions}
We have established the map from a Grassmann functional $G [\psi,\hat\psi ]$ to a family of probability distributions of a classical statistical ensemble $\{p_{\tau}(t) \}$ at different times $t$. In particular, this map also specifies the unitary time evolution of $\{p_{\tau}(t) \}$. The existence of this map is very general -- it only requires a relation between the ``boundary terms'' $\hat g_f$ and $g_{in}$ in eq. \eqref{M1B} and a suitable form of the action $S$. Our examples cover two-state quantum systems as well as quantum field theories for fermions. ``Diagonal observables'' find a direct interpretation in terms of the classical probabilities. Such diagonal observables are realized as diagonal quantum operators acting on the real ``classical wave function'' $q_{\tau}=s_{\tau}\sqrt{p_{\tau}}$, $s_{\tau}=\pm 1$, according to
\be \label{conc1}
(\hat A q)_{\tau}=\sum_{\rho} A_{\tau\rho}q_{\rho}, \qquad A_{\tau\rho}=A_{\tau}\delta_{\tau\rho}.
\ee
They obey the standard classical rule for expectation values
\be \label{conc2}
\langle A \rangle=\sum_{\tau} p_{\tau} A_{\tau},
\ee
where $A_{\tau}$ is associated to the value of $A$ in the state $\tau$. Eq. \eqref{conc2} is  equivalent to the  ``quantum rule''
\be \label{conc3}
\begin{split}
\langle A\rangle &=\langle q \hat A q\rangle=\sum_{\tau,\rho} q_{\tau}A_{\tau\rho}q_{\rho} \\
&=\sum_{\tau}q_{\tau}^2 A_{\tau}=\sum_{\tau} p_{\tau} A_{\tau}.
\end{split}
\ee
We also can express $\langle A \rangle$ as a standard functional integral
\be \label{conc4}
\langle A \rangle= \int {\cal D}\psi {\cal D}\hat\psi  A[\hat\psi ,\psi] G[\psi,\hat\psi ],
\ee
with $A[\hat\psi,\psi ]$ a Grassmann element associated to the observable.

In general, the map $G\to \{p_{\tau}(t) \}$ is not invertible. While we can establish an invertible map between $G$ and the family of real wave functions $\{q_{\tau}(t) \}$, we observe that two wave functions differing only by signs $s_{\tau}$ will yield the same probability distribution $\{p_{\tau}(t) \}$. This ``gauge ambiguity'' of the choice of the sign function $s_{\tau}(t)$ does not affect the expectation values of diagonal observables \eqref{conc3}. Furthermore, the choice of $s_{\tau}(t)$ can be largely fixed by requirements of continuity (and differentiability) of $\{q_{\tau}(t) \}$ in time (and space if appropriate). The residual gauge freedom can be fixed by a suitable convention which does not influence the prediction of expectation values. After gauge fixing, we can construct the inverse map $\{p_{\tau}(t) \}\to \{q_{\tau}(t) \}$, which allows us to compute the wave function $\{q_{\tau} \}$ for a given probability density $\{p_{\tau} \}$. For a suitable time evolution of $\{p_{\tau}(t) \}$ we can map our classical statistical ensemble to the quantum formalism with wave function $\{q_{\tau}(t) \}$. In turn, this can be mapped to a Grassmann functional $G [\psi,\hat\psi ]$ and the associated functional integral. In this way we can explicitly construct a quantum field theory of fermions from a classical statistical ensemble.

If the time evolution of the real wave function $\{q_\tau(t)\}$ is compatible with a suitable involution defining a complex structure it can be mapped to the usual complex wave function of quantum mechanics. We have discussed several examples for such complex structures. In particular, the quantum field theory for two-dimensional fermions admits a complex structure associated to the notion of antiparticles, with involution corresponding to charge conjugation. We emphasize that different complex structures, based on different involutions are possible as well. For example, complex structures can be associated to discrete symmetries arising for quantum field theories with different ``flavors'' of fermionic fields.

The construction of a quantum field theory for fermions from a classical probability distribution is not limited to two dimensions. For example, we may generalize the action \eqref{D24} to four dimensions, where a real representation of the four matrices $\gamma^\mu$ as $4\times 4$ matrices exists. This involves instead of $\psi_\pm(x)$ four species of Grassmann variables $\psi_\gamma(x)$. A real one particle wave function describes a Majorana spinor with four components. In four dimensions, Majorana spinors are equivalent to Weyl spinors \cite{CWMS}. Extension to a complex Grassmann algebra similar to sect. \ref{qft} accounts for Dirac spinors. One further may consider several flavors of fermions and add mass terms and interactions. 

Once a quantum field theory for fermions is obtained from a classical statistical ensemble, we may consider quantum states of a single particle or of several particles. Particles can be interpreted here as isolated excitations of a vacuum state. Their properties depend on the choice of the vacuum state. We emphasize in this context that the state $|0\kr$ in eq. \eqref{D9} is not the unique possibility for a vacuum state. One could equally well consider the charge conjugate state ${\cal C}|0\kr$ or a linear combination of $|0\kr$ and ${\cal C}|0\kr$. Another interesting possibility is the half filled state with an equal number of particles and antiparticles or holes. 

For the description of a particle state much less information is needed as compared to the whole classical probability distribution $\{p_\tau\}$. This is rather obvious, since $\{p_\tau\}$ contains additional information for all multi-particle states. An isolated subsystem as a one particle state is therefore characterized by a coarse graining of the information. For our two dimensional example with vacuum $|0\kr$ all one-particle observables must be computable from the information contained in the wave function $q_{1,0}(x)$ \eqref{D8} or, more generally, from the associated density matrix. This generalizes to arbitrary isolated one particle excitations. 

In presence of a complex structure the complex one-particle wave function $\psi_Q^{(1)}(x)$ in quantum mechanics is typically constructed from two real ``classical'' wave functions $q_1(x)$ and $q_2(x),\psi^{(1)}_Q(x)=q_1(x)+iq_2(x)$, cf. eq. \eqref{S14}. The quantum mechanical probability density $p_Q(x)=|\psi^{(1)}_Q(x)|^2=q^2_1(x)+q^2_2(x)$ is not enough to describe the physics of interference. In our case, however, we dispose of two separate ``coarse grained'' probability distributions $p_1(x)=q^2_1(x)$ and $p_2(x)=q^2(x)$. Together with sign-information for $q_1$ and $q_2$ this constitutes sufficient information for the description of phases and interference phenomena. Derivative operators as the quantum mechanical momentum can be associated \cite{CW??} to suitable classical observables $A_\tau$ of the ``microscopic classical ensemble'' with states $\tau$.

Often the classical correlation functions for the one-particle-observables can no longer be computed from the one-particle density matrix. In this case we have to deal with ``incomplete statistics'' \cite{3}. The correlations between measurements of different one-particle-observables have to be described by new types of correlation functions \cite{GR}. These new types of correlation functions can be implemented within the classical statistical ensemble, based on product structures for observables that differ from the classical product. In this way non-commutativity arises from the coarse graining of the information \cite{CW??}. The necessary use of a non-commuting product for the description of measurement correlations in isolated subsystems explains \cite{GR,CW??} why those violate Bell's inequalities \cite{Bell}. In this way the ``no go theorems'' \cite{BS,KS} for an implementation of quantum mechanics within classical statistics are circumvented \cite{GR}. Our examples of fermionic quantum field theories constitute indeed explicit realizations of the emergence of quantum physics from classical statistics.

\end{document}